\documentclass[aps, prd, letterpaper, 12pt, nofootinbib, superscriptaddress, floatfix, longbibliography]{revtex4-2}

\usepackage[utf8]{inputenc}
\usepackage[dvips]{graphics}
\usepackage[usenames,dvipsnames]{color}
\usepackage{amsmath,amsfonts,amssymb,amsthm}
\usepackage{mathrsfs}
\usepackage{graphicx} 
\usepackage[compat=1.1.0]{tikz-feynman}
\usepackage{subcaption}

\makeatletter
\let\p@section\@empty
\let\p@subsection\@empty
\let\p@subsubsection\@empty
\makeatother

\definecolor{hgreen}{rgb}{0,.7,0}
\definecolor{hred}{rgb}{.7,0,0}
\definecolor{hblue}{rgb}{0,0,.7}

\usepackage[colorlinks=true,
linkcolor=hblue,
citecolor=hgreen,
filecolor=hblue,
urlcolor=hred]{hyperref}

\allowdisplaybreaks

\usepackage[many]{tcolorbox}
\usepackage{tikz}

\begin{document}
   
\title{Targets for Flavor-Violating Top Decays}

\author{Wolfgang~Altmannshofer}
\email{waltmann@ucsc.edu}
\affiliation{Department of Physics, University of California Santa Cruz, and
Santa Cruz Institute for Particle Physics, 1156 High St., Santa Cruz, CA 95064, USA}

\author{Zev~Balme}
\email{zbalme@ucsc.edu}
\affiliation{Department of Physics, University of California Santa Cruz, and
Santa Cruz Institute for Particle Physics, 1156 High St., Santa Cruz, CA 95064, USA}

\author{Christopher~M.~Donohue}
\email{cmdonohu@ucsc.edu}
\affiliation{Department of Physics, University of California Santa Cruz, and
Santa Cruz Institute for Particle Physics, 1156 High St., Santa Cruz, CA 95064, USA}

\author{Stefania~Gori}
\email{sgori@ucsc.edu}
\affiliation{Department of Physics, University of California Santa Cruz, and
Santa Cruz Institute for Particle Physics, 1156 High St., Santa Cruz, CA 95064, USA}

\author{Siddharth~Vignesh~Mukundhan}
\email{smukundh@ucsc.edu}
\affiliation{Department of Physics, University of California Santa Cruz, and
Santa Cruz Institute for Particle Physics, 1156 High St., Santa Cruz, CA 95064, USA}

\begin{abstract}
Analyticity and unitarity constrain certain classes of new physics models by linking flavor-conserving and flavor-violating four-fermion interactions. In this work, we explore how these theoretical relations impact flavor-violating rare top quark decays. Building on our previous results, we present an updated analysis of the decays $t \to q \ell^+ \ell^-$ and identify interesting target branching ratios in the range of $10^{-7}$ to $10^{-6}$ once current experimental constraints from flavor-conserving processes are taken into account. We extend the analysis to top decays with lepton flavor violation, deriving correlations among the relevant Wilson coefficients and confronting them with existing limits from LEP and the LHC. Notably, we find that current searches for $t \to q e \mu$ are already probing theoretically motivated regions of parameter space. These results strongly support continued efforts to explore flavor-violating top decays as a powerful probe of new physics.
\end{abstract}

\maketitle
\newpage
\tableofcontents
\newpage

\section{Introduction} \label{sec:intro}

The top quark plays a crucial role in electroweak symmetry breaking and serves as a natural probe for new physics beyond the Standard Model (BSM). Despite extensive studies of its production and decay properties at the Large Hadron Collider (LHC), many of its characteristics remain unexplored, particularly in the context of flavor-changing neutral currents (FCNCs). In the SM, the FCNC decays of the type $t\to q X$ (with $X=h,Z,\gamma,g$) are highly suppressed due to the Glashow-Iliopoulos-Maiani (GIM) mechanism, with branching ratios predicted to be well below $\mathcal O(10^{-10})$~\cite{Aguilar-Saavedra:2004mfd, Altmannshofer:2019ogm, Balaji:2020qjg}, several orders of magnitude below the current LHC limits and the expected ultimate LHC reach~\cite{ATLAS:2024kxj}. Therefore, the observation of flavor-changing top decays would be an unambiguous sign of new physics. Many UV models can predict the enhancement of these branching ratios at levels that are accessible in experiment. 

If new physics is too heavy to be produced on-shell, it can still leave experimental imprints that can be described by the framework of an effective field theory (EFT) (see e.g.~\cite{Drobnak:2008br, Durieux:2014xla, Forslund:2018qcp, Chala:2018agk, Shi:2019epw, Afik:2021jjh, Altmannshofer:2023bfk} for the study of top FCNCs in the EFT context and~\cite{Bahl:2023xkw, Cremer:2023gne, Bradshaw:2023wco, Frank:2023fkc, Jueid:2024cge} for other recent work on top FCNCs). Interestingly, the Wilson coefficients that appear in the EFT Lagrangian are constrained not only by data, but also by theoretical arguments based on unitarity, locality, and analyticity \cite{Adams:2006sv}. 

The authors of~\cite{Remmen:2020uze} applied these arguments to an EFT that contains four-fermion operators finding that the Wilson coefficients are subject to ``sum rule'' relations (see also~\cite{Bellazzini:2016xrt, Remmen:2020vts, Gu:2020thj, Davighi:2021osh, Azatov:2021ygj, Zhang:2021eeo, Remmen:2022orj} for related work). In particular, certain flavor-violating dimension six four-fermion operators are upper bounded by combinations of the Wilson coefficients of flavor-conserving operators. The bounds are valid under the following assumptions: (1) the UV scale of the higher-dimensional operators far exceeds SM fermion masses, (2) the UV theory is Lorentz invariant, unitary, and local; (3) the UV theory sufficiently improves the scaling of the forward four-fermion scattering amplitude at large Mandelstam variable, $s$, in order to avoid poles at infinity; (4) the UV theory is dominated by either scalars or vectors~\cite{Remmen:2020uze}.\footnote{\label{foot:models} It is worth pointing out that these assumptions are fairly restrictive. Known UV models that obey the assumptions are scalar and vector leptoquarks. However, assumption (3) is violated in models in which flavor-changing semileptonic operators are mediated by $Z^\prime$ gauge bosons. Assumption (4) is generically violated in models in which the four-fermion operators are loop-induced. In such models, the bounds on flavor-violating dimension six four-fermion operators can be violated.}

In~\cite{Altmannshofer:2023bfk}, we showed that combining such theoretical constraints on the Wilson coefficients with experimental limits from flavor-conserving processes leads to predictions for the branching ratio of rare three-body top decays, $t\to q\ell^+\ell^-$, that are approximately an order of magnitude below current experimental sensitivities across the class of new physics models that obey the assumptions made in~\cite{Remmen:2020uze}. As a result, any future observation of such flavor-violating top decays would necessarily imply a violation of the sum rules relations. The primary goal of this paper is to extend our analysis to include both quark and lepton flavor-violating operators mediating the decays $t \to q \ell\ell^\prime$, with $\ell,\ell^\prime =e,\mu$, $\ell\neq\ell^\prime$. In this context, we first present the sum rule relations for the relevant four-fermion operators and then analyze their phenomenological implications. Notably, we find that the theoretical upper bounds on BR$(t \to q \ell\ell^\prime)$ are comparable to the current LHC bounds. This underscores the importance of ongoing and future searches for lepton and quark flavor-violating top decays in the coming years.

This paper is organized as follows: in Sec.~\ref{sec:positivity}, we introduce the EFT setup for the rare top decays and present the necessary theoretical conditions the quark and lepton flavor-violating Wilson coefficients need to satisfy. In Sec.~\ref{sec:observables}, we discuss the most relevant experimental constraints coming from low energy searches for lepton flavor-violating processes, LHC searches for rare top decays, LEP searches for single top production, LHC measurements of single top, $t\bar t$, and di-lepton production, precision $Z$ boson decay rate measurements, and rare $B$ decays. A summary of the main findings of this section is presented in Figure~\ref{fig:chart} as well as in Sec.~\ref{sec:summary}. In Sec.~\ref{sec:numerics}, we present our numerical analysis and discuss the impact of the theoretical relations on lepton flavor-violating and lepton flavor-conserving rare top quark decays. We reserve Sec.~\ref{sec:conclusions} for our conclusions.

\section{Positivity Constraints on Flavor-Changing Top Quark Interactions} \label{sec:positivity}

Our analysis is based on an effective Lagrangian framework featuring four-fermion operators relevant for rare top decays into semileptonic final states, specifically $t \to q \ell^+ \ell^-$ and $t \to q \ell \ell^\prime$, where $q = u, c$, $\ell,\ell^\prime=e,\mu, \ell\neq\ell^\prime$. Consequently, we focus on semileptonic operators involving up-type quarks and charged leptons, particularly those subject to the sum-rule conditions established in~\cite{Remmen:2020uze}.
We further restrict our analysis to operators containing right-handed up-type quarks. This is because flavor-changing interactions involving left-handed up-type quarks are generically linked via the $SU(2)_L$ symmetry to operators that mediate flavor-changing transitions among left-handed down-type quarks, which are tightly constrained by experimental data on rare $B$ meson decays~\cite{Fox:2007in}. Finally, we do not consider operators with tau leptons, due to the better prospects for searching for final states with muons and electrons. The relevant part of the effective Lagrangian we consider is thus
\begin{multline}
 \mathcal L \supset \frac{1}{\Lambda^2} \sum_{\ell = e,\mu} \sum_{q = u,c,t} \Big[ C^{LR}_{\ell \ell qq} (\bar \ell \gamma^\alpha P_L \ell) + C^{RR}_{\ell \ell qq} (\bar \ell \gamma^\alpha P_R \ell) \Big] (\bar q \gamma_\alpha P_R q) \\ 
 + \frac{1}{\Lambda^2} \sum_{\ell = e,\mu} \sum_{q = u,c} \Big[ C^{LR}_{\ell \ell t q} (\bar \ell \gamma^\alpha P_L \ell) + C^{RR}_{\ell \ell t q} (\bar \ell \gamma^\alpha P_R \ell) \Big] (\bar t \gamma_\alpha P_R q) ~+~ \text{h.c.} \\
 + \frac{1}{\Lambda^2} \sum_{q = u,c,t} \Big[ C^{LR}_{e \mu qq} (\bar e \gamma^\alpha P_L \mu) + C^{RR}_{e \mu qq} (\bar e \gamma^\alpha P_R \mu) \Big] (\bar q \gamma_\alpha P_R q) ~+~ \text{h.c.} \\\label{eq:L}
 + \frac{1}{\Lambda^2} \sum_{q = u,c} \Big[ C^{LR}_{ e \mu t q} (\bar e \gamma^\alpha P_L \mu) + C^{RR}_{e \mu t q} (\bar e \gamma^\alpha P_R \mu) + C^{LR}_{ \mu e t q} (\bar \mu \gamma^\alpha P_L e) \\ + C^{RR}_{\mu e t q} (\bar \mu \gamma^\alpha P_R e) \Big] (\bar t \gamma_\alpha P_R q) ~+~ \text{h.c.} ~.
\end{multline}
The terms in the Lagrangian can be categorized as follows:
\begin{itemize}
\item The first line contains $\Delta F = 0$ operators, which preserve flavor. The corresponding Wilson coefficients are $C^{LR}_{\ell\ell qq}$ and $C^{RR}_{\ell\ell qq}$.
\item The second and third lines correspond to $\Delta F = 1$ operators, where flavor changes occur either in the lepton or quark sector. The corresponding Wilson coefficients are $C^{LR}_{\ell \ell tq}$, $C^{RR}_{\ell \ell tq}$ or $C^{LR}_{e \mu qq}$, $C^{RR}_{e \mu qq}$, respectively.
\item The fourth and fifth lines represent $\Delta F = 2$ operators, where flavor transitions occur in both sectors. Those Wilson coefficients are $C^{LR}_{e \mu tq}$, $C^{RR}_{e \mu tq}$, $C^{LR}_{\mu e tq}$, and $C^{RR}_{\mu e tq}$. 
\end{itemize}
Note that our notation for the Wilson coefficients follows~\cite{Altmannshofer:2023bfk} which slightly differs from the commonly used SMEFT notation~\cite{Grzadkowski:2010es, Aguilar-Saavedra:2018ksv}.

Using S-matrix analyticity and partial wave unitarity, one can derive sum-rule relations that constrain the Wilson coefficients~\cite{Remmen:2020uze}. In particular, under the assumptions detailed in~\cite{Remmen:2020uze} and mentioned in the introduction, the signs of the $\Delta F = 0$ Wilson coefficients are fixed. If the UV theory that gives rise to (\ref{eq:L}) is dominated by scalars, then one finds
\begin{equation}
 C_{\ell\ell qq}^{LR} < 0 ~, \quad  C_{\ell\ell qq}^{RR} > 0 ~.
\end{equation}
If the UV is dominated by vectors, then one instead finds
\begin{equation}
 C_{\ell\ell qq}^{LR} > 0 ~, \quad  C_{\ell\ell qq}^{RR} < 0 ~.
\end{equation}
In both cases, one finds that the size of the $\Delta F = 1$ coefficients is limited by the size of the $\Delta F = 0$ coefficients~\cite{Remmen:2020uze, Altmannshofer:2023bfk} 
\begin{equation} \label{eq:sum_rule_DeltaF1}
    \left|C^{XY}_{\ell \ell^\prime qq}\right|  \leq \sqrt{C^{XY}_{\ell\ell qq} C^{XY}_{\ell^\prime \ell^\prime qq} } ~,\qquad \left|C^{XY}_{\ell \ell qq^\prime}\right|  \leq \sqrt{C^{XY}_{\ell\ell qq} C^{XY}_{\ell\ell q^\prime q^\prime} }  ~, \quad \text{with}~ XY = LR, RR ~.
\end{equation}
If the scalar and vector contributions are of similar magnitude, they may cancel each other out, and no bounds on the flavor-changing coefficients can be established.

We now describe a procedure to obtain bounds on the $\Delta F = 2$ coefficients. To the best of our knowledge, such bounds have not been reported in the literature. Our starting point is an inequality established in~\cite{Remmen:2020uze}
\begin{equation} \label{eq:sumrule}
\pm C^{XY}_{ijkl} \alpha_i \alpha_j^* \beta_k^* \beta_l > 0~, \end{equation} 
where $\alpha$ and $\beta$ are arbitrary vectors in flavor space. The sign is determined by the chirality indicated by $X$ and $Y$, as well as whether the UV contributions to the Wilson coefficients are scalar or vector in nature \cite{Remmen:2020uze}. The power of the inequality lies in the fact that it has to hold for all choices of the flavor vectors $\alpha$ and $\beta$. 

For the positive case in equation~\eqref{eq:sumrule} (analogous reasoning applies to the negative case), we introduce the following parameterization of the vectors $\alpha$ and $\beta$
\begin{equation}
\alpha_i = e^{i\phi_\alpha} \sin{\theta_\alpha} \delta_{i\ell} + \cos{\theta_\alpha} \delta_{i \ell^\prime} ~, \quad
\beta_i = e^{i\phi_\beta} \sin{\theta_\beta} \delta_{i q} + \cos{\theta_\beta} \delta_{i q^\prime} ~,
\end{equation}
with the assumptions $\ell \neq \ell^\prime$ and $q \neq q^\prime$. Without loss of generality, we take both $\sin\theta_i$ and $\cos\theta_i$ to be positive.
Note that these parameterizations are not fully generic; they only take into account two flavors of quarks and two flavors of leptons. The bounds that we will find using these parameterizations are therefore necessary but not sufficient. Substituting these forms into the inequality~\eqref{eq:sumrule} and rewriting the coefficients as $C^{XY}_{ijkl} = |C^{XY}_{ijkl}| e^{i\phi_{ijkl}}$, we can isolate the magnitudes $|C^{XY}_{\ell \ell^\prime q q^\prime}|$ and $|C^{XY}_{\ell^\prime \ell q q^\prime}|$ by making the choice 
\begin{equation}
\cos(\phi_\alpha - \phi_\beta + \phi_{\ell \ell^\prime q q^\prime}) = \cos( \phi_\alpha + \phi_\beta - \phi_{\ell^\prime \ell q q^\prime}) = -1~,
\end{equation}
which fixes the phases $\phi_\alpha$ and $\phi_\beta$ up to a simultaneous shift $\phi_\alpha \to \phi_\alpha \pm \pi$ and $\phi_\beta \to \phi_\beta \pm \pi$. Introducing the notation $t_\alpha = \tan\theta_\alpha$ and $t_\beta = \tan\theta_\beta$, we learn that
\begin{multline} \label{eq:cosines}
\left|C^{XY}_{\ell \ell^\prime q q^\prime}\right| + \left|C^{XY}_{\ell^\prime \ell q q^\prime}\right| \leq \frac{1}{2} \left( t_\alpha t_\beta C^{XY}_{\ell\ell qq}  +  \frac{t_\alpha}{t_\beta} C^{XY}_{\ell\ell q^\prime q^\prime} + \frac{t_\beta}{t_\alpha} C^{XY}_{\ell^\prime \ell^\prime qq} + \frac{1}{t_\alpha t_\beta} C^{XY}_{\ell^\prime \ell^\prime q^\prime q^\prime}  \right) \\
+ t_\alpha \left|C^{XY}_{\ell\ell qq^\prime}\right| \cos{(\phi_{\ell\ell qq^\prime} - \phi_\beta)}  + \frac{1}{t_\alpha} \left|C^{XY}_{\ell^\prime \ell^\prime qq^\prime}\right| \cos{(\phi_{\ell^\prime \ell^\prime qq^\prime} - \phi_\beta)} \\
+ t_\beta \left|C^{XY}_{\ell\ell^\prime qq}\right| \cos{(\phi_{\ell\ell^\prime qq} + \phi_\alpha)} + \frac{1}{t_\beta}\left|C^{XY}_{\ell\ell^\prime q^\prime q^\prime}\right| \cos{(\phi_{\ell\ell^\prime q^\prime q^\prime} + \phi_\alpha)} ~.
\end{multline}
The four terms in the second and third line all depend linearly on cosines that contain either $\phi_\alpha$ or $\phi_\beta$. Using the freedom to shift both of these phases by $\pm \pi$, the sum of the four terms can be made either positive or equal in magnitude but negative. Summing the two relations in equation \eqref{eq:cosines}, with $(\phi_\alpha, \phi_\beta)$ and $(\phi_\alpha \pm \pi, \phi_\beta \pm \pi)$, we learn that
\begin{equation}
\left|C^{XY}_{\ell \ell^\prime q q^\prime}\right| + \left|C^{XY}_{\ell^\prime \ell q q^\prime}\right| \leq \frac{1}{2} \left( t_\alpha t_\beta C^{XY}_{\ell\ell qq} + \frac{t_\alpha}{t_\beta} C^{XY}_{\ell \ell q^\prime q^\prime}  + \frac{t_\beta}{t_\alpha} C^{XY}_{\ell^\prime \ell^\prime qq} + \frac{1}{t_\alpha t_\beta} C^{XY}_{\ell^\prime \ell^\prime q^\prime q^\prime} \right) ~.
\end{equation}
The optimal bound is obtained by setting
\begin{equation}
\tan\theta_\alpha = \sqrt[4]{\frac{C^{XY}_{\ell^\prime \ell^\prime qq} C^{XY}_{\ell^\prime \ell^\prime q^\prime q^\prime}}{C^{XY}_{\ell\ell qq} C^{XY}_{\ell\ell q^\prime q^\prime}}}~, \quad \tan\theta_\beta = \sqrt[4]{\frac{C^{XY}_{\ell\ell q^\prime q^\prime } C^{XY}_{\ell^\prime \ell^\prime q^\prime q^\prime}}{C^{XY}_{\ell\ell qq} C^{XY}_{\ell^\prime \ell^\prime qq}}} ~,
\end{equation}
and reads
\begin{equation}  \label{eq:sum_rule_DeltaF2}
    \left|C^{XY}_{\ell \ell^\prime q q^\prime}\right| + \left|C^{XY}_{\ell^\prime \ell q q^\prime}\right| \leq \sqrt{C^{XY}_{\ell \ell qq} C^{XY}_{\ell^\prime \ell^\prime q^\prime q^\prime} }+\sqrt{ C^{XY}_{\ell \ell q^\prime q^\prime} C^{XY}_{\ell^\prime \ell^\prime qq}} ~.
\end{equation}
The same bound applies whether the UV is dominated by scalars of vectors.
Given the various choices we made throughout the derivation, equation~\eqref{eq:sum_rule_DeltaF2} is only a necessary but not a sufficient condition. In particular, stronger bounds may apply if one considers sets of Wilson coefficients that involve all three flavors of quarks and leptons simultaneously. 

Our goal is to explore the impact of the relations in equations~\eqref{eq:sum_rule_DeltaF1} and \eqref{eq:sum_rule_DeltaF2} in view of the existing experimental information we have on the relevant Wilson coefficients.

As already mentioned in footnote~\ref{foot:models}, we reiterate here that the relations are only valid in certain UV models, in particular scalar or vector leptoquark models, but not in $Z^\prime$ models or models with loop-induced flavor changing operators. In the context of flavor-changing semileptonic top decays, the relevant leptoquarks are (in the notation of~\cite{Dorsner:2016wpm})
\begin{eqnarray}
\text{scalar:} &&\quad R_2 \sim (\mathbf{3},\mathbf{2},7/6) ~, \quad S_1 \sim (\mathbf{\bar 3}, \mathbf{1}, 1/3) ~, \\
\text{vector:} &&\quad \tilde V_2 \sim (\mathbf{\bar 3},\mathbf{2},-1/6) ~, \quad \tilde U_1 \sim (\mathbf{3}, \mathbf{1}, 5/3) ~.
\end{eqnarray}
It is straightforward to show that UV models with a single leptoquark saturate the relations in~\eqref{eq:sum_rule_DeltaF1} and~\eqref{eq:sum_rule_DeltaF2}. In models with multiple scalar or vector leptoquarks, the inequalities in the relations~\eqref{eq:sum_rule_DeltaF1} and~\eqref{eq:sum_rule_DeltaF2} are a consequence of the Cauchy-Schwarz inequality. Formulating the relations in an EFT setup allows us to capture the physics of all UV scenarios with a single or multiple leptoquarks that are either scalars or vectors.

\section{Experimental Observables} \label{sec:observables}

In this section, we collect the most important experimental probes that are sensitive to the various Wilson coefficients that enter the sum-rule relations discussed above. 

\begin{figure}
    \centering
    \includegraphics[width=1.0\textwidth]{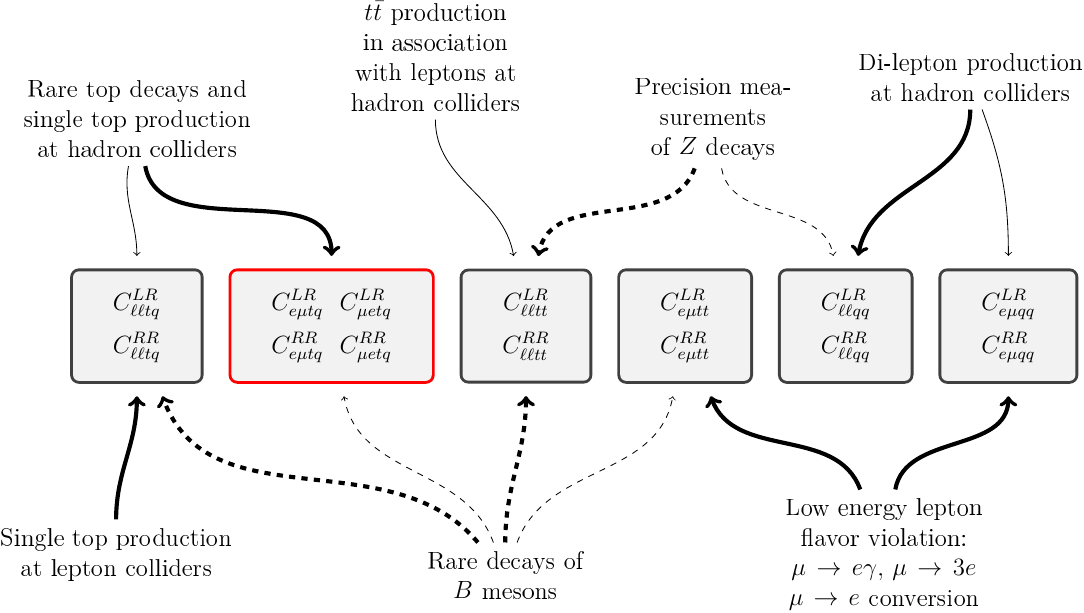}
    \caption{The Wilson coefficients of the four-fermion operators considered in this work and their most relevant experimental probes. Thick arrows indicate observables that have the highest sensitivity. Rare $B$ and $Z$ decays are affected by the considered Wilson coefficients at the loop level and are thus less robust (indicated by the dashed arrows). The main novelty of our paper is the analysis of the $\Delta F =2$ coefficients indicated in red.}
    \label{fig:chart}
\end{figure}

The chart in Figure~\ref{fig:chart} summarizes the full set of processes we consider and illustrates the best probes of each class of Wilson coefficients. Thick arrows indicate the observables that have the highest sensitivity on each Wilson coefficient. Dashed arrows indicate loop level constraints which are less robust than those that arise at tree level (solid lines). The coefficients highlighted in red are analyzed in the context of EFT sum rule constraints for the first time in our paper.

In the following subsections we consider low energy lepton flavor-violating processes (section~\ref{sec:lfv}), rare top decays as well as single top production at the LHC (section~\ref{sec:raretop}), single top production at lepton colliders (section~\ref{sec:ee_to_tq}), rare decays of $B$ mesons (section~\ref{sec:Bdecay}), di-lepton production at the LHC (section~\ref{sec:DY}), flavor-conserving decays of the $Z$ boson (section~\ref{sec:Zdecay}), and top quark pair production in association with leptons at the LHC (section~\ref{sec:ttbar}).

\subsection{Low energy lepton flavor-changing processes} \label{sec:lfv}

Lepton flavor-violating and quark flavor-conserving four-fermion operators are strongly constrained by low energy searches for lepton flavor-violating processes, $\mu$ to $e$ conversion in nuclei in particular. In the past, the SINDRUM experiment set a bound on the lepton flavor-violating conversion rate utilizing gold nuclei, $\Gamma(\mu^- {\rm Au}\to e^- {\rm Au})/\Gamma(\mu^- {\rm Au})<7\times 10^{-13}$ at 90$\%$ C.L.~\cite{SINDRUMII:2006dvw}, where $\Gamma(\mu^- {\rm Au})$ denotes the nuclear muon capture rate. In the future, the Mu2e~\cite{Mu2e:2014fns, Bernstein:2019fyh} and COMET~\cite{COMET:2018auw, Moritsu:2022lem} experiments are expected to improve the sensitivity by roughly four orders of magnitude.

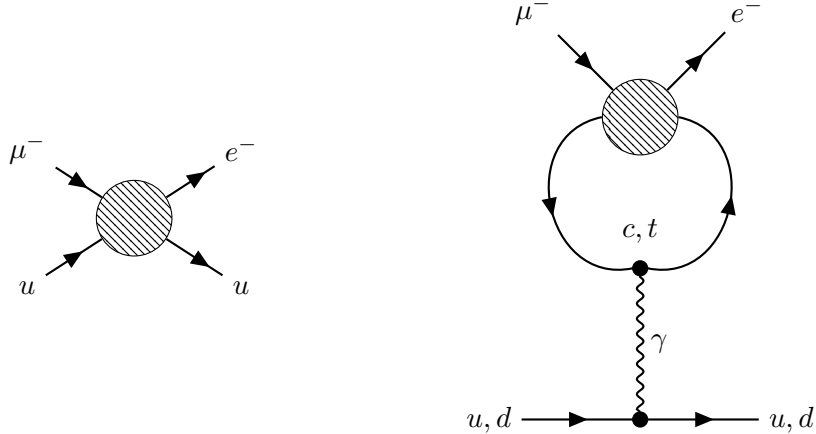
\begin{figure}[tb]
\centering
    \begin{minipage}{0.4\textwidth}
        \centering
        \begin{tikzpicture} 
        \begin{feynman}[large]
            \vertex (f1) {\(\mu^-\)}; 
            \node [below right=of f1, yshift=.5cm, style=blob] (a); 
            \vertex [below left=of a, yshift=.5cm](f2) {\(u\)}; 
            \vertex [above right=of a, yshift=-.5cm] (f3) {\(e^{-}\)};
            \vertex [below right=of a, yshift=.5cm] (f4) {\({u}\)};
            
            \diagram* {
                (f1) -- [fermion] (a), 
                (f2) -- [fermion] (a),  
                (f3) -- [anti fermion] (a),
                (f4) -- [anti fermion] (a),
            }; 
        \end{feynman}
    \end{tikzpicture}
    \end{minipage}
    \begin{minipage}{0.4\textwidth}
         \centering
         \begin{tikzpicture} 
             \begin{feynman}[large]
                 \vertex (i1) {\(\mu^-\)};
                 \node [below right=of i1, style=blob] (A);
                 \vertex [above right=of A] (f1) {\(e^{-}\)};
                 \node [below=of A, style=dot] (B);
                 \node [below=of B, style=dot] (C);
                 \vertex [left=of C] (i2) {\(u,d\)};
                 \vertex [right=of C] (f2) {\(u,d\)};

                 \node at ($(A) + (0,-1.5)$) {\(c,t\)};
        
                 \diagram*{
                 (i1) -- [fermion] (A) -- [fermion] (f1),
                 (A) -- [fermion, half right] (B),
                 (B) -- [fermion, half right] (A),
                 (B) -- [photon, edge label=\(\gamma\)] (C),
                 (i2) -- [fermion] (C),
                 (C) -- [fermion] (f2)
                 };
             \end{feynman}
         \end{tikzpicture}
     \end{minipage}
\caption{Example Feynman diagrams contributing to $\mu \to e$ conversion in nuclei at tree level (left) and at the 1-loop level (right). The blobs correspond to the effective four-fermion interactions.}
\label{fig:mu2e}
\end{figure}

The quark flavor-conserving Wilson coefficients $C_{e\mu uu}^{LR}$ and $C_{e\mu uu}^{RR}$ contribute at the tree level to $\mu$ to $e$ conversion. Similarly, at the one loop level, the corresponding coefficients with top and charm quarks contribute. Example Feynman diagrams are shown in Figure~\ref{fig:mu2e}.

The SINDRUM bound leads to a constraint on the Wilson coefficients of the lepton flavor-violating operators involving top quarks at the level of~\cite{Garosi:2023yxg}
\begin{equation}
\frac{\left| C_{e\mu tt}^{LR} \right|}{\Lambda^2} , \frac{\left| C_{e\mu tt}^{RR} \right|}{\Lambda^2}\lesssim  \frac{1}{( 100 \,\text{TeV})^2} ~.
\end{equation}
The corresponding constraints on the Wilson coefficients with up or charm quarks are at least as stringent. Future bounds from Mu2e and COMET will probe even smaller Wilson coefficients at the level of $\Lambda \gtrsim 10^3$\,TeV to $10^4$\,TeV depending on the operator~\cite{Haxton:2024lyc} (see also~\cite{Davidson:2020hkf}). These scales are much larger than the ones of the other operators we will discuss in this paper.

Lepton flavor-violating operators can, in principle, also be constrained by other low-energy processes, like the $\mu\to e\gamma$ decay~\cite{MEGII:2023ltw} and the $\mu\to 3e$ decay~\cite{SINDRUM:1987nra}, as well as by LHC searches for non-standard $e\mu$ production~\cite{Angelescu:2020uug, ATLAS:2020tre, CMS:2022fsw}. However, these constraints are much weaker than the one coming from $\mu\to e$ conversion, and we do not consider them in the following. 
Lepton flavor-violating operators can also be constrained by $B$ decays. However, as we will argue in section \ref{sec:Bdecay}, these constraints are again much weaker than the constraints from $\mu\to e$ transitions.

We also checked if quark and lepton flavor violating operators can be probed by $\mu \to e$ conversion. In fact, the Wilson coefficients $C_{\mu e t q}^{LR}$, $C_{\mu e t q}^{RR}$, $C_{e \mu t q}^{LR}$, and $C_{e \mu t q}^{RR}$ contribute through 1-loop $W$ exchange to $(\mu e)(dd)$ operators and thus can be constrained using limits on $\mu \to e$ conversion rates\footnote{We thank Lorenzo Calibbi for pointing this out.}. Although such loop contributions to $\mu \to e$ conversion are suppressed by small CKM elements and quark masses, we find interesting constraints. Using the results from~\cite{Davidson:2020hkf} we find for the quark flavor-violating Wilson coefficients with right-handed charm quarks
\begin{equation}
\frac{|C_{\mu e t c}^{LR}|}{\Lambda^2}, \frac{|C_{\mu e t c}^{RR}|}{\Lambda^2}, \frac{|C_{e \mu t c}^{LR}|}{\Lambda^2}, \frac{|C_{e \mu t c}^{RR}|}{\Lambda^2} \lesssim \frac{1}{( 200\,\text{GeV})^2} ~,
\end{equation}
only slightly weaker than the bounds from $B$ decays discussed in section~\ref{sec:Bdecay}. For the quark flavor-violating Wilson coefficients with right-handed up quarks, we do not find meaningful bounds from $\mu \to e$ conversion.

We conclude that because of the stringent constraints from $\mu\to e$ conversion, the Wilson coefficients $C_{e\mu qq}^{LR}$, $C_{e\mu qq}^{RR}$, $C_{e\mu tt}^{LR}$, and $C_{e\mu tt}^{RR}$ are not relevant for our analysis and we will largely neglect them in the following sections. 

\subsection{Rare top decays and single top production in association with leptons in proton-proton collisions} \label{sec:raretop}

\begin{figure}[tb]
\centering
    \begin{minipage}{0.4\textwidth}
        \centering
        \begin{tikzpicture} 
            \begin{feynman}[large]
                \vertex (a) {\(t\)}; 
                \node [right=of a, style=blob] (b); 
                \vertex [above right=of b](f1) {\(\ell^{-}\)}; 
                \vertex [right=of b] (f2) {\(\ell^{+}\)};
                \vertex [below right=of b] (f3) {\(q\)};
                
                \diagram* {
                    (a) -- [fermion] (b), 
                    (b) -- [fermion] (f1),  
                    (b) -- [anti fermion] (f2),
                    (b) -- [fermion] (f3),
                }; 
            \end{feynman}
        \end{tikzpicture}
    \end{minipage}
    \begin{minipage}{0.4\textwidth}
        \centering
        \begin{tikzpicture} 
            \begin{feynman}[large]
                \vertex (a) {\(q\)}; 
                \node [right=of a, style=blob] (b); 
                \vertex [right=of b,yshift=.5cm](f1) {\(\ell^{-}\)}; 
                \vertex [right=of b,yshift=-.5cm] (f2) {\(\ell^{+}\)};
                \vertex[below=of a] (g) {\(g\)};
                \node [below=of b,style=dot] (c);
                \vertex [right=of c] (f3) {\(t\)};
                
                \diagram* {
                    (a) -- [fermion] (b), 
                    (b) -- [fermion] (f1),  
                    (b) -- [anti fermion] (f2),
                    (b) --[fermion, edge label=\(t\)] (c),
                    (g) -- [gluon] (c),
                    (c) -- [fermion] (f3)
                }; 
            \end{feynman}
        \end{tikzpicture}
    \end{minipage}
\caption{Example Feynman diagrams contributing to rare semileptonic top decays (left) and single top production in association with leptons in proton-proton collisions (right). The blobs correspond to the effective four-fermion interactions. The diagrams show lepton flavor-conserving processes. Lepton flavor-violating versions are also possible.}
\label{fig:t_decay}
\end{figure}
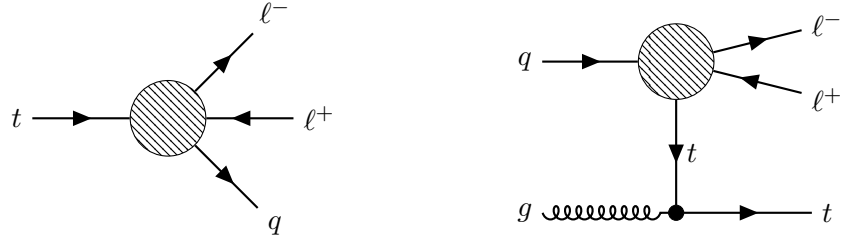

Rare semileptonic top decays and single top production in association with leptons in proton-proton collisions are directly related processes and probe the same new physics interactions~\cite{Durieux:2014xla, Davidson:2015zza}. The lepton flavor-conserving processes $t \to q \ell^+ \ell^-$ and $pp \to t \ell^+ \ell^-$, where $q = u,c$ and $\ell = e, \mu$ are sensitive to the effective couplings $C_{\ell \ell tq}^{LR}$ and $C_{\ell \ell tq}^{RR}$, while the lepton flavor-changing processes $t \to q \ell \ell^\prime$ and $pp \to t \ell \ell^\prime$ with $\ell \neq \ell^\prime$, constrain the lepton flavor-violating couplings $C_{\ell \ell^\prime tq}^{LR}$ and $C_{\ell \ell^\prime tq}^{RR}$.
Example Feynman diagrams are shown in Figure~\ref{fig:t_decay}.

On the experimental side, search results for the top decays and for single top production can be combined to optimize the sensitivity. This has been done in the lepton flavor-violating case in the CMS searches~\cite{CMS:2022ztx, CMS:2023iul} and the results are conveniently quoted in terms of top branching ratios. The most stringent constraints relevant for our analysis read at 95\% C.L.~\cite{CMS:2023iul} 
\begin{equation} \label{eq:bounds_tqllp}
\text{BR}(t \to u e \mu) < 2.2\times 10^{-8} ~,\quad \text{BR}(t \to c e \mu) < 3.7 \times 10^{-7} ~,
\end{equation}
where the quoted branching ratio bounds refer to the sum of the two possible leptonic modes $\text{BR}(t \to q e \mu) = \text{BR}(t \to q e^+ \mu^-) + \text{BR}(t \to q e^- \mu^+)$.
Similar searches have been performed for processes involving tau leptons~\cite{ATLAS:2024njy}, but they are beyond the scope of this work. The constraints are stronger for up quarks than for charm quarks, due to the larger up-quark PDFs of the proton. 

In the presence of the lepton flavor-violating effective interactions, one finds to a good approximation
\begin{multline}
\text{BR}(t \to q \ell \ell^\prime) \simeq \frac{1}{96 \pi^2} \frac{m_t^2 v^2}{\Lambda^4} \left( \bigl| C^{LR}_{\ell\ell^\prime t q} \bigr|^2 + \bigl| C^{RR}_{\ell\ell^\prime t q} \bigr|^2  + \bigl| C^{LR}_{\ell^\prime\ell t q} \bigr|^2 + \bigl| C^{RR}_{\ell^\prime\ell t q} \bigr|^2 \right) \\
\times \left(1 - \frac{m_W^2}{m_t^2}\right)^{-2} \left(1 + \frac{2m_W^2}{m_t^2}\right)^{-1} ~.
\end{multline}
where $v \simeq 246$\,GeV is the Higgs vacuum expectation value and $m_t \simeq 172.4$\,GeV~\cite{ParticleDataGroup:2022pth} is the top pole mass.
From the limits in equation~\eqref{eq:bounds_tqllp} we find the following constraints on the combination of the Wilson coefficients and the new physics scale $\Lambda$
\begin{eqnarray}
  \frac{|C_{e\mu tu}^{LR}|}{\Lambda^2} , ~\frac{|C_{e\mu tu}^{RR}|}{\Lambda^2} ,~ \frac{|C_{\mu etu}^{LR}|}{\Lambda^2} , ~\frac{|C_{\mu etu}^{RR}|}{\Lambda^2}  &<& \frac{1}{(3.1\,  \text{TeV})^2} ~, \label{eq:tuemu_bound} \\
  \frac{|C_{e\mu tc}^{LR}|}{\Lambda^2} , ~\frac{|C_{e\mu tc}^{RR}|}{\Lambda^2} ,~ \frac{|C_{\mu etc}^{LR}|}{\Lambda^2} , ~\frac{|C_{\mu etc}^{RR}|}{\Lambda^2}  &<& \frac{1}{(1.6\,  \text{TeV})^2} ~.  \label{eq:tcemu_bound}
\end{eqnarray}
These limits were obtained with 138\,fb$^{-1}$ of data~\cite{CMS:2023iul}. Scaling to the 3\,ab$^{-1}$ luminosity expected at the high-luminosity (HL) LHC, we estimate that the sensitivity to the branching ratios might improve by a factor of 5, resulting in a sensitivity to an approximately 50\% higher new physics scale.

To the best of our knowledge, no dedicated LHC searches exist for the lepton flavor-conserving three-body decays $t \to q \ell^+ \ell^-$, or the corresponding single top production in association with leptons. In~\cite{Chala:2018agk}, constraints on the top decay branching ratios have been obtained by recasting a run II ATLAS search for the decays $t \to Z q$~\cite{ATLAS:2018zsq}. Translating the constraints into our framework, we find at 95\% C.L.~\cite{Altmannshofer:2023bfk}
\begin{eqnarray}
  \text{BR}(t \to c e^+ e^-) < 2.1 \times 10^{-4} ~, && \quad  \text{BR}(t \to c \mu^+ \mu^-) < 1.5 \times 10^{-4} ~, \\
  \text{BR}(t \to u e^+ e^-) < 1.8 \times 10^{-4} ~, && \quad \text{BR}(t \to u \mu^+ \mu^-) < 1.2 \times 10^{-4} ~.
\end{eqnarray}

To a good approximation, the corresponding theory expressions for the branching ratios are given by 
\begin{equation}
\text{BR}(t \to q \ell^+ \ell^-) \simeq \frac{1}{96 \pi^2} \frac{m_t^2 v^2}{\Lambda^4} \left(\bigl| C^{LR}_{\ell\ell t q} \bigr|^2 + \bigl| C^{RR}_{\ell\ell t q} \bigr|^2 \right) \left( 1 - \frac{m_W^2}{m_t^2}\right)^{-2} \left(1 + \frac{2m_W^2}{m_t^2} \right)^{-1} ~.
\end{equation}
With the current observed limits on the branching ratios, one finds fairly weak constraints on the corresponding new physics scale 
\begin{eqnarray} \label{eq:C_raretop1}
  \frac{|C_{eetc}^{LR}|}{\Lambda^2} , ~ \frac{|C_{eetc}^{RR}|}{\Lambda^2} < \frac{1}{(0.32\, \text{TeV})^2} ~, && \quad \frac{|C_{eetu}^{LR}|}{\Lambda^2} , ~ \frac{|C_{eetu}^{RR}|}{\Lambda^2} < \frac{1}{(0.33\, \text{TeV})^2}  ~, \\  \label{eq:C_raretop2}
  \quad \frac{|C_{\mu\mu tc}^{LR}|}{\Lambda^2} , ~ \frac{|C_{\mu\mu tc}^{RR}|}{\Lambda^2} < \frac{1}{(0.35\, \text{TeV})^2} ~, && \quad \frac{|C_{\mu\mu tu}^{LR}|}{\Lambda^2} , ~ \frac{|C_{\mu\mu tu}^{RR}|}{\Lambda^2} < \frac{1}{(0.36\, \text{TeV})^2}  ~.
\end{eqnarray}
While an EFT description of the new physics can in principle be valid for scales as low as the top mass, one can expect sizable corrections of the order of $m_t^2/\Lambda^2$ in concrete new physics models.

Reference~\cite{Chala:2018agk} estimates that dedicated searches at the HL-LHC might improve the sensitivity to branching ratios at the $ \sim 10^{-6}$ level, corresponding to new physics scales of around $1$\,TeV (see also~\cite{Bostanabad:2025zyb}). Further improvement by another order of magnitude in branching ratio sensitivity would be possible at a 100 TeV proton-proton collider~\cite{FCC:2018byv}.

\subsection{Single top production at lepton colliders} \label{sec:ee_to_tq}

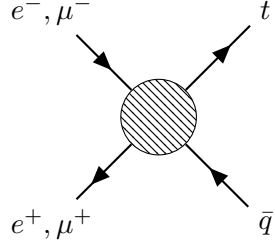
\begin{figure}[tb]
    \centering
    \begin{tikzpicture} 
        \begin{feynman}[large]
            \vertex (f1) {\(e^{-}, \mu^-\)}; 
            \node [below right=of f1,style=blob] (a); 
            \vertex [below left=of a](f2) {\(e^{+}, \mu^+\)}; 
            \vertex [above right=of a] (f3) {\(t\)};
            \vertex [below right=of a] (f4) {\(\bar{q}\)};
            
            \diagram* {
                (f1) -- [fermion] (a), 
                (f2) -- [anti fermion] (a),  
                (f3) -- [anti fermion] (a),
                (f4) -- [fermion] (a),
            }; 
        \end{feynman}
    \end{tikzpicture}
    \caption{Example Feynman diagram contributing to single top production at lepton colliders.}
    \label{fig:single_t}
\end{figure}

An alternative probe of top flavor-changing and lepton flavor-conserving interactions is single top production at lepton colliders. An example diagram is shown in Figure~\ref{fig:single_t}.

Introducing a CP-averaged cross section in electron-positron collisions
\begin{equation}
    \sigma(e^+ e^- \to t q) = \sigma(e^+ e^- \to t \bar q) + \sigma(e^+ e^- \to \bar t q)~,
\end{equation}
one finds the following expression in terms of the Wilson coefficients
\begin{equation} \label{eq:single-top}
    \sigma(e^+ e^- \to t q) = \frac{1}{6\pi} \frac{s}{\Lambda^4} \left(1-\frac{m_t^2}{s}\right)^2\left( 1 + \frac{m_t^2}{2s} \right) \left( \big| C^{LR}_{eetq}\big|^2 + \big| C^{RR}_{eetq}\big|^2 \right)~,
\end{equation}
where $s$ is the center of mass energy squared, and $m_t \simeq 172.4$~\cite{ParticleDataGroup:2022pth} is the top pole mass.
An analogous expression also holds for the cross section of $\mu^+ \mu^- \to tq$ at a muon collider. Above we have assumed that the lepton beams are unpolarized. 

Constraints on single top production have been obtained at LEP~\cite{Aleph:2001dzz}. As shown in~\cite{Durieux:2014xla}, the best bound on the Wilson coefficients can be obtained from the search at a center of mass energy of $\sqrt{s} = 207$\,GeV, which resulted in the following upper limit at $95\%$ C.L.
\begin{equation}
\sigma(e^+ e^- \to t q)_\text{LEP} < 0.17 ~ \text{pb}  \quad \text{for} ~~ \sqrt{s} = 207~\text{GeV}~.
\end{equation}
As charm tagging was not applied in the LEP search, we interpret the constraint as a constraint on the sum of up and charm production
\begin{equation}
\sigma(e^+ e^- \to t q) = \sigma(e^+ e^- \to t u) + \sigma(e^+ e^- \to t c) ~.
\end{equation}
The corresponding constraints on the new physics scale are almost a factor of three stronger than the one obtained from the rare top decays, c.f. equations~\eqref{eq:C_raretop1} and~\eqref{eq:C_raretop2},
\begin{equation} \label{eq:bounds_single_top}
  \frac{|C_{eetu}^{LR}|}{\Lambda^2} , ~ \frac{|C_{eetu}^{RR}|}{\Lambda^2} ,~ \frac{|C_{eetc}^{LR}|}{\Lambda^2} , ~ \frac{|C_{eetc}^{RR}|}{\Lambda^2} < \frac{1}{(0.9\, \text{TeV})^2} ~.
\end{equation}

One can expect much improved sensitivity to single top production at future $e^+ e^-$ colliders, such as FCC-ee, CEPC, and ILC~\cite{Bernardi:2022hny, ILCInternationalDevelopmentTeam:2022izu, CEPCPhysicsStudyGroup:2022uwl, Ai:2024nmn}. 
The authors of~\cite{Shi:2019epw} estimate that new physics scales of several TeV could be probed at the Higgs factory run of the CEPC. 
We expect that similar sensitivities could be reached at FCC-ee and at the ILC.

We also note that there are excellent prospects for probing the $(\mu\mu)(t q)$ operators at a future muon collider. Searches for single top production in $\mu^+ \mu^-$ collisions at a center of mass energy of $\sqrt{s} = 10$\,TeV could give sensitivities to the Wilson coefficients $C_{\mu\mu tq}^{XY}$ that approach a level of $\sim 100$~TeV~\cite{Sun:2023cuf,  Bhattacharya:2023beo,  Ake:2023xcz}. 

\subsection{\texorpdfstring{\boldmath}{} Rare decays of \texorpdfstring{$B$}{B} mesons} \label{sec:Bdecay}

\begin{figure}[tb]
    \centering
\begin{tikzpicture}
        \begin{feynman}[large]
            \vertex (B) {\(b\)};
            \node [right=of B,style=dot] (a);
            \node[right=of a, style=blob] (c);
            \vertex [right=of c,yshift=.5cm] (f1) {\(\ell^{-}\)};
            \vertex [right=of c,yshift=-.5cm] (f2) {\(\ell^{+}\)};
            \node [above right=of c,style=dot] (d);
            \vertex [above right=of d] (S) {\(s\)};
    
            \diagram* {
                (B) -- [fermion] (a),
                (a) -- [fermion,edge label=\(t\)] (c),
                (c) -- [fermion, edge label=\(t\,c\)] (d),
                (d) -- [boson, quarter right, edge label=\(W\)] (a),
                (c) -- [fermion] (f1),
                (c) -- [anti fermion] (f2),
                (d) -- [fermion] (S)
            };
        \end{feynman}
    \end{tikzpicture}
    \caption{Example Feynman diagram contributing to lepton flavor-conserving rare decays of $B$ mesons based on the quark level $b \to s \ell^+ \ell^-$ transition. Analogous contributions to lepton flavor-violating rare $B$ decays also exist.}
    \label{fig:B_decay}
\end{figure}
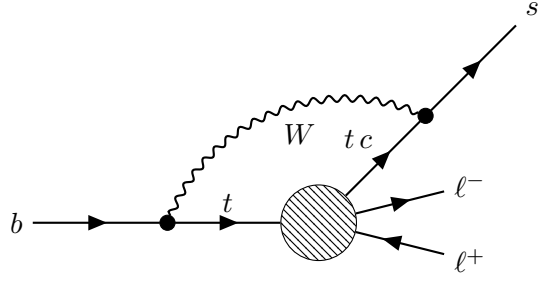

Effective quark flavor-conserving and flavor-violating top quark interactions can contribute to rare $b \to s \ell^+ \ell^-$ and $b \to s \ell \ell^\prime$ decays at the 1-loop level, see e.g.~\cite{Aebischer:2015fzz, Celis:2017doq, Camargo-Molina:2018cwu, Bissmann:2019gfc, Bissmann:2020mfi, Bruggisser:2021duo, Grunwald:2023nli, Garosi:2023yxg}. In the following, we discuss both the lepton flavor-conserving decays $b \to s \mu^+ \mu^-$ and $b \to s e^+ e^-$ and the lepton flavor-violating decays $b \to s e \mu$. An example diagram for a lepton flavor-conserving decay is shown in Figure~\ref{fig:B_decay}.

\paragraph{Lepton flavor-conserving rare $B$ decays decays:}
We closely follow the analysis in~\cite{Altmannshofer:2023bfk}. 
The rare $B$ decays of interest are governed by an effective Hamiltonian 
\begin{equation}\label{eq:HB}
 \mathcal H_\text{eff} = - \frac{4 G_F}{\sqrt{2}} V_{tb} V_{ts}^* \frac{\alpha}{4 \pi} \sum_\ell \Big( C_9^\ell O_9^\ell + C_{10}^\ell O_{10}^\ell + ~\text{h.c.}  \Big) ~,
\end{equation}
with the dimension 6 operators
\begin{equation}\label{eq:HB1}
O_9^\ell = (\bar s \gamma_\alpha P_L b)(\bar \ell \gamma^\alpha \ell) ~,\qquad O_{10}^\ell = (\bar s \gamma_\alpha P_L b)(\bar \ell \gamma^\alpha \gamma_5 \ell) ~,
\end{equation}
and the corresponding Wilson coefficients at the leading log approximation are
\begin{equation} \label{eq:C9}
    C_9^\ell = \left( C^{RR}_{\ell\ell tt} + C^{LR}_{\ell\ell tt} + \frac{V_{cs}^*}{V_{ts}^*} \frac{m_c}{m_t} \big(C^{RR}_{\ell\ell tc} + C^{LR}_{\ell\ell tc} \big)^* \right) \frac{1}{4 s_W^2} \frac{m_t^2}{4m_W^2} \frac{v^2}{\Lambda^2} \log\left(\frac{\Lambda^2}{m_W^2}\right)~,
\end{equation}
\begin{equation} \label{eq:C10}
    C_{10}^\ell = \left( C^{RR}_{\ell\ell tt} - C^{LR}_{\ell\ell tt} + \frac{V_{cs}^*}{V_{ts}^*}\frac{m_c}{m_t} \big(C^{RR}_{\ell\ell tc} - C^{LR}_{\ell\ell tc} \big)^* \right) \frac{1}{4 s_W^2} \frac{m_t^2}{4m_W^2} \frac{v^2}{\Lambda^2}\log\left(\frac{\Lambda^2}{m_W^2}\right) ~.
\end{equation}
In these expressions $m_t \simeq 162.9$\,GeV is the top quark $\overline{\text{MS}}$ mass and $s_W^2 \simeq 0.231$ is the sine squared of the weak mixing angle~\cite{ParticleDataGroup:2024cfk}. For our numerical analysis we use $m_c / m_t \simeq 3.69 \times 10^{-3}$ (based on $\overline{\text{MS}}$ values evaluated at a common scale) and approximate $V_{cs}^* / V_{ts}^* \simeq -24.3$ (based on tree level CKM input and ignoring the small imaginary part of $V_{ts}$). As in~\cite{Altmannshofer:2023bfk}, we only include the logarithmically enhanced terms. Additional finite corrections are of the same order as unknown model dependent UV matching contributions to the $B$ decays. The constraints we derive from $B$ decays thus rely on the assumption that the log terms are dominant. If that is not the case, concrete new physics models might give $\mathcal O(1)$ corrections to our results.

Global fits of rare $B$ decay data result in bounds on the new physics contributions to the Wilson coefficients $C_9^\ell$ and $C_{10}^\ell$~(see e.g.~\cite{Altmannshofer:2021qrr, SinghChundawat:2022zdf, Ciuchini:2022wbq, Greljo:2022jac, Alguero:2023jeh, Wen:2023pfq, Altmannshofer:2023uci, Guadagnoli:2023ddc, Hurth:2023jwr, Bordone:2024hui, Fleischer:2025ucq} for recent work). As we are interested in finding constraints on new physics, we focus our analysis on the theoretically clean lepton flavor universality tests~\cite{Belle:2016fev, BELLE:2019xld, Belle:2019oag, LHCb:2021lvy, LHCb:2022qnv, LHCb:2022vje, CMS:2024syx, LHCb:2024rto, LHCb:2024yci, LHCb:2025pxz} and the $B_s \to \mu^+ \mu^-$ branching ratio~\cite{ATLAS:2018cur, LHCb:2021awg, CMS:2022mgd}. We update the results from~\cite{Altmannshofer:2023bfk} by including the CMS measurements of $R_K$ and $R_{K^*}$~\cite{CMS:2024syx}\footnote{We have not included the recent LHCb results on lepton flavor universality in $B_s \to \phi \ell^+ \ell^-$ decays~\cite{LHCb:2024rto}, $B^+ \to K^+\pi^+\pi^-\ell^+\ell^-$ decays~\cite{LHCb:2024yci} or $B^0 \to K^{*\,0} \ell^+ \ell^-$ angular observables~\cite{LHCb:2025pxz}. Given the uncertainties in these measurements, we do not expect these results to have a significant impact on our analysis.} and find
\begin{equation}
C_{10}^\mu = -0.011 \pm 0.228 ~,~~~ C_{10}^e = 0.469 \pm 0.683 ~, ~~~ C_9^\mu - C_9^e = -0.708 \pm 0.687 ~,
\end{equation}
with the error correlation matrix 
\begin{equation}
\rho = \begin{pmatrix} 1 & 0.293 & 0.104 \\ 0.293 & 1 & - 0.897 \\ 0.104 & -0.897 & 1 \end{pmatrix} ~. 
\end{equation}
With our approach, the direction $C_9^\mu = C_9^e$ remains unconstrained. 

We illustrate the constraints that can be obtained from the rare $B$ decays by setting a single coefficient to $\pm 1$ and deriving limits on the corresponding new physics scale using the criterion $\Delta \chi^2 < 4$. We find results similar to the ones in~\cite{Altmannshofer:2023bfk}
\begin{align} \label{eq:bound_B_tt_LR}
  \frac{-1}{(3.2\, \text{TeV})^2} < \frac{C_{eett}^{LR}}{\Lambda^2} &< \frac{1}{(1.2\, \text{TeV})^2} ~,& \quad \frac{-1}{(1.3\, \text{TeV})^2} < \frac{C_{\mu\mu tt}^{LR}}{\Lambda^2} &< \frac{1}{(3.3\, \text{TeV})^2} ~, \\  \label{eq:bound_B_tt_RR}
  \frac{-1}{(0.53\, \text{TeV})^2} < \frac{C_{eett}^{RR}}{\Lambda^2} &< \frac{1}{(0.34\, \text{TeV})^2} ~,& \quad \frac{-1}{(0.93\, \text{TeV})^2} < \frac{C_{\mu\mu tt}^{RR}}{\Lambda^2} &< \frac{1}{(0.83\, \text{TeV})^2} ~, \\\label{eq:bound_B_tq_LR}
  \frac{-1}{(0.22\, \text{TeV})^2} < \frac{C_{ee tc}^{LR}}{\Lambda^2} &< \frac{1}{(0.74\, \text{TeV})^2} ~,& \quad \frac{-1}{(0.76\, \text{TeV})^2} < \frac{C_{\mu\mu tc}^{LR}}{\Lambda^2} &< \frac{1}{(0.25\, \text{TeV})^2} ~.  
  \end{align}
We do not obtain meaningful limits in the cases of $C_{eetc}^{RR}$ and $C_{\mu\mu tc}^{RR}$.

With the data that will be collected during the high-luminosity phase of the LHC, one can expect that the reach to high new physics scales can improve by a factor of 2-3, see e.g.~\cite{Altmannshofer:2023uci}.  

\bigskip
\paragraph{Lepton flavor-violating rare $B$ decays decays:}
Of particular interest are the decays $B_s \to \mu e$, $B^+ \to K^+ \mu e$, $B^0 \to K^{*0} \mu e$, and $B_s \to \phi \mu e$ for which stringent upper limits exist from LHCb~\cite{LHCb:2017hag, LHCb:2019bix, LHCb:2022lrd}. Searches for lepton flavor-violating $B$ decays with taus in the final state also exist~\cite{LHCb:2022wrs}, but they are beyond the scope of our work.

In the following, we list the relevant expressions for the branching ratios in terms of lepton flavor-violating Wilson coefficients $C_9^{\ell\ell^\prime}$ and $C_{10}^{\ell\ell^\prime}$, defined similarly as in equations (\ref{eq:HB}), (\ref{eq:HB1}) (see for example~\cite{Becirevic:2016zri, Becirevic:2024vwy}). The expressions for these Wilson coefficients are analogous to the expressions for lepton flavor-conserving $B$ decays discussed above 
\begin{equation} \label{eq:C9_llp}
    C_9^{\ell \ell^\prime} = \left( C^{RR}_{\ell\ell^\prime tt} + C^{LR}_{\ell\ell^\prime tt} + \frac{V_{cs}^*}{V_{ts}^*} \frac{m_c}{m_t} \big(C^{RR}_{\ell\ell^\prime tc} + C^{LR}_{\ell\ell^\prime tc} \big)^* \right) \frac{1}{4 s_W^2} \frac{m_t^2}{4m_W^2} \frac{v^2}{\Lambda^2} \log\left(\frac{\Lambda^2}{m_W^2}\right)~,
\end{equation}
\begin{equation} \label{eq:C10_llp}
    C_{10}^{\ell \ell^\prime} = \left( C^{RR}_{\ell\ell^\prime tt} - C^{LR}_{\ell\ell^\prime tt} + \frac{V_{cs}^*}{V_{ts}^*}\frac{m_c}{m_t} \big(C^{RR}_{\ell\ell^\prime tc} - C^{LR}_{\ell\ell^\prime tc} \big)^* \right) \frac{1}{4 s_W^2} \frac{m_t^2}{4m_W^2} \frac{v^2}{\Lambda^2}\log\left(\frac{\Lambda^2}{m_W^2}\right) ~.
\end{equation}

\noindent
For the $B_s \to \mu e$ branching ratio we have 
\begin{equation} \label{eq:Bsmue}
\text{BR}(B_s \to \mu e) \simeq \tau_{B_s} \frac{G_F^2 \alpha^2}{64 \pi^3} f_{B_s}^2 m_{B_s} m_\mu^2 |V_{tb} V_{ts}^*|^2 \Big( |C_9^{\mu e}|^2 + |C_9^{e\mu}|^2 + |C_{10}^{\mu e}|^2 + |C_{10}^{e \mu}|^2 \Big) ~,
\end{equation}
where we have neglected the lepton masses, which is an excellent approximation. 
In the above expression, $\tau_{B_s}$ is the lifetime of the heavy $B_s$ mass eigenstate, $\tau_{B_s} = (1.622 \pm 0.008)$\,ps~\cite{ParticleDataGroup:2024cfk}. The given expression for the branching ratio refers to the sum of $\mu^+ e^-$ and $\mu^- e^+$ decay modes $\text{BR}(B_s \to \mu e) = \text{BR}(B_s \to \mu^+ e^-) + \text{BR}(B_s \to \mu^- e^+) $, and we use the same notation for the semileptonic decays below. 
The analytic expressions for the $B_s \to \mu^+ e^-$ and $B_s \to \mu^- e^+$ branching ratios separately contain only the $C_{9,10}^{e\mu}$ or $C_{9,10}^{\mu e}$ coefficients, respectively (and the analogous statement holds for the semileptonic decays below). For the $B_s$ meson decay constant that enters equation~\eqref{eq:Bsmue} we use $f_{B_s} = (230.3 \pm 1.3)$\,MeV~\cite{Dowdall:2013tga, ETM:2016nbo, Bazavov:2017lyh, Hughes:2017spc}.

For the semileptonic decay $B^+ \to K^+ \mu e$ we find
\begin{multline}
\text{BR}(B^+ \to K^+ \mu e) \simeq \tau_{B^+} \frac{G_F^2 \alpha^2}{1536 \pi^5} m_{B^+}^3 |V_{tb} V_{ts}^*|^2 \Big( |C_9^{\mu e}|^2 + |C_9^{e\mu}|^2 + |C_{10}^{\mu e}|^2 + |C_{10}^{e \mu}|^2 \Big) \\
\times \int_0^{q^2_\text{max}} dq^2 ~ \lambda^\frac{3}{2}\left( 1 , \frac{m_{K^+}^2}{m_{B^+}^2} , \frac{q^2}{m_{B^+}^2} \right) f_+^2(q^2)~,
\end{multline}
with $q^2_\text{max} = (m_{B^+} - m_{K^+} )^2$ and the Kallen function is $\lambda(a,b,c) = a^2 + b^2 + c^2 - 2(ab + bc + ca)$. We use the $B^+ \to K^+$ form factor $f_+$ from~\cite{Gubernari:2023puw} which is based on the BSZ parameterization~\cite{Bharucha:2015bzk}.

Finally, the corresponding expression for $B^0 \to K^{*0} \mu e$ is
\begin{multline}
\text{BR}(B^0 \to K^{*0} \mu e) \simeq \tau_{B^0} \frac{G_F^2 \alpha^2}{768 \pi^5} m_{B^0}^3 |V_{tb} V_{ts}^*|^2 \Big( |C_9^{\mu e}|^2 + |C_9^{e\mu}|^2 + |C_{10}^{\mu e}|^2 + |C_{10}^{e \mu}|^2 \Big) \\
\times \int_0^{q^2_\text{max}} dq^2 ~ \lambda^\frac{1}{2}\left( 1 , \frac{m_{K^{*0}}^2}{m_{B^0}^2} , \frac{q^2}{m_{B^0}^2} \right) \left[ 32 A_{12}^2(q^2) \frac{m_{K^{*0}}^2}{m_{B^0}^2} + A_1^2(q^2) \frac{q^2}{m_{B^0}^2} \left( 1 + \frac{m_{K^{*0}}}{m_{B^0}}\right)^2 \right. \\
\left. + V^2(q^2) \frac{q^2}{m_{B^0}^2} \left( 1 + \frac{m_{K^{*0}}}{m_{B^0}}\right)^{-2}  \lambda\left( 1 , \frac{m_{K^{*0}}^2}{m_{B^0}^2} , \frac{q^2}{m_{B^0}^2} \right)\right] ~.
\end{multline}
Here, the maximal $q^2$ value is given by $q^2_\text{max} = (m_{B^0} - m_{K^{*0}} )^2$. We use the $B^0 \to K^{*0}$ form factors $A_{12}$, $A_1$ and $V$ from~\cite{Gubernari:2023puw} based on the BSZ parameterization~\cite{Bharucha:2015bzk}.

The expression for $B_s \to \phi \mu e$ is completely analogous to $B^0 \to K^{*0} \mu e$ with the obvious replacements of the meson masses, lifetimes, and the form factors.

Experimentally, the upper bounds from LHCb are (see also~\cite{Belle:2018peg} for a previous Belle search for $B^0 \to K^{*0} \mu e$)
\begin{eqnarray}
\text{BR}(B_s \to \mu e) &<& 6.3 \times 10^{-9} ~, \quad \text{\cite{LHCb:2017hag}} \\
\text{BR}(B^+ \to K^+ \mu^- e^+) &<& 9.5 \times 10^{-9} ~, \quad \text{\cite{LHCb:2019bix}} \\
\text{BR}(B^+ \to K^+ \mu^+ e^-) &<& 8.8 \times 10^{-9} ~, \quad \text{\cite{LHCb:2019bix}} \\
\text{BR}(B^0 \to K^{*0} \mu^- e^+) &<& 7.9 \times 10^{-9} ~, \quad \text{\cite{LHCb:2022lrd}} \\
\text{BR}(B^0 \to K^{*0} \mu^+ e^-) &<& 6.9 \times 10^{-9} ~, \quad \text{\cite{LHCb:2022lrd}} \\
\text{BR}(B^0 \to K^{*0} \mu e) &<& 11.7 \times 10^{-9} ~, \quad \text{\cite{LHCb:2022lrd}} \\
\text{BR}(B_s \to \phi \mu e) &<& 19.8 \times 10^{-9} ~, \quad \text{\cite{LHCb:2022lrd}}
\end{eqnarray}
with all limits at the $95\%$ confidence level. Note that the quoted limit on $B_s \to \mu e$ assumes that it is the heavy $B_s$ mass eigenstate that decays into $\mu e$~\cite{LHCb:2017hag}, as is the case for our new physics scenario.

We translate the experimental limits into constraints on the new physics scale by switching on one coefficient at a time and considering the strongest of the constraints from the above list of branching ratio bounds. We find
\begin{eqnarray} \label{eq:C_rareB1}
  \frac{|C_{e\mu tt}^{LR}|}{\Lambda^2} , ~ \frac{|C_{e\mu tt}^{RR}|}{\Lambda^2} &<& \frac{1}{(1.0\, \text{TeV})^2} ~,  \\  \label{eq:C_rareB2}
  \frac{|C_{e\mu tc}^{LR}|}{\Lambda^2} , ~ \frac{|C_{e\mu tc}^{RR}|}{\Lambda^2} &<& \frac{1}{(0.30\, \text{TeV})^2}  ~, \\ 
  \label{eq:C_rareB3}
  \frac{|C_{\mu e tc}^{LR}|}{\Lambda^2} , ~ \frac{|C_{\mu e tc}^{RR}|}{\Lambda^2} &<& \frac{1}{(0.29\, \text{TeV})^2}  ~.
\end{eqnarray}
In all cases, the strongest constraint is obtained from the $B^0 \to K^{* 0} \mu^+ e^-$ or $B^0 \to K^{* 0} \mu^- e^+$ decay, respectively.

Based on the expected increase in statistics at the HL-LHC, the sensitivity to the branching ratios can improve by more than an order of magnitude. The corresponding sensitivity to the new physics scale might increase approximately by a factor of 2.  

As anticipated in section~\ref{sec:lfv}, the sensitivities to the top flavor-conserving coefficients $C_{e\mu tt}^{LR}$ and $C_{e\mu tt}^{RR}$ from $\mu$ to $e$ conversion are much stronger than the ones from the rare $B$ decays in~\eqref{eq:C_rareB1}. 
We also observe that the rare $B$ decay sensitivity to the top flavor-violating coefficients is considerably weaker than the sensitivity of LHC searches for rare top decays found in equation~\eqref{eq:tcemu_bound}.

\subsection{Di-lepton production at the LHC} \label{sec:DY}

\begin{figure}[tb]
    \centering
\begin{tikzpicture} 
        \begin{feynman}[large]
            \vertex (f1) {\(q\)}; 
            \node [below right=of f1,style=blob] (a); 
            \vertex [below left=of a](f2) {\(\bar{q}\)}; 
            \vertex [above right=of a] (f3) {\(\ell^{-}\)};
            \vertex [below right=of a] (f4) {\(\ell^{+}\)};
            
            \diagram* {
                (f1) -- [fermion] (a), 
                (f2) -- [anti fermion] (a),  
                (f3) -- [anti fermion] (a),
                (f4) -- [fermion] (a),
            }; 
        \end{feynman}
    \end{tikzpicture}
    \caption{Example Feynman diagram contributing to di-lepton production in proton-proton collisions.}
    \label{fig:dilepton}
\end{figure}
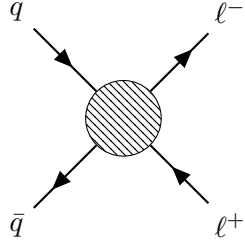

Di-lepton production at the LHC, $pp \to \ell^+ \ell^-$, constitutes a well-known probe of flavor-conserving new physics contact interactions of quarks and leptons (see e.g.~\cite{Hiller:2025hpf} for a recent study). A relevant Feynman diagram is shown in Figure~\ref{fig:dilepton}. Similarly to~\cite{Altmannshofer:2023bfk}, we use the ATLAS analysis of $pp \to \ell^+ \ell^-$ in~\cite{Aad:2020otl} to find constraints on the Wilson coefficients. The corresponding CMS analysis~\cite{CMS:2021ctt} would result in similar constraints.

We calculate the $pp \to \ell^+ \ell^-$ cross sections with the \texttt{high-PT} package~\cite{Allwicher:2022gkm, Allwicher:2022mcg}, using \texttt{NNPDF3.1} parton distributions functions~\cite{NNPDF:2017mvq}.
Following~\cite{Altmannshofer:2023bfk}, we calculate the expected number of new physics events in the signal regions of the ATLAS analysis using
\begin{equation}
    N_\text{sig}^\ell \simeq N_\text{bg}^\ell \left( \frac{\sigma(pp \to \ell^+\ell^-)}{\sigma(pp \to \ell^+\ell^-)_\text{SM}} - 1 \right) ~,
\end{equation}
where $N_\text{bg}^\ell$ are the expected number of SM events in the various signal regions. To derive the constraints on the Wilson coefficients, we impose the 95\% C.L. upper limits on the number of new physics signal events, $N_\text{sig}^\ell$, given in~\cite{Aad:2020otl}.

Switching on one operator at a time, we collect here the constraints on the new physics scale that we find
\begin{align} \label{eq:bounds_ppll_1}
  \frac{-1}{(7.5\, \text{TeV})^2} < \frac{C_{eeuu}^{LR}}{\Lambda^2} &< \frac{1}{(6.2\, \text{TeV})^2} ~,& \quad \frac{-1}{(8.3\, \text{TeV})^2} < \frac{C_{eeuu}^{RR}}{\Lambda^2} &< \frac{1}{(5.6\, \text{TeV})^2} ~, \\ \label{eq:bounds_ppll_2}
  \frac{-1}{(2.4\, \text{TeV})^2} < \frac{C_{eecc}^{LR}}{\Lambda^2} &< \frac{1}{(2.3\, \text{TeV})^2} ~,& \quad \frac{-1}{(2.4\, \text{TeV})^2} < \frac{C_{eecc}^{RR}}{\Lambda^2} &< \frac{1}{(2.3\, \text{TeV})^2} ~, \\ \label{eq:bounds_ppll_3}
  \frac{-1}{(8.1\, \text{TeV})^2} < \frac{C_{\mu\mu uu}^{LR}}{\Lambda^2} &< \frac{1}{(5.9\, \text{TeV})^2} ~,& \quad \frac{-1}{(9.3\, \text{TeV})^2} < \frac{C_{\mu\mu uu}^{RR}}{\Lambda^2} &< \frac{1}{(5.1\, \text{TeV})^2} ~, \\ \label{eq:bounds_ppll_4}
  \frac{-1}{(2.3\, \text{TeV})^2} < \frac{C_{\mu\mu cc}^{LR}}{\Lambda^2} &< \frac{1}{(2.3\, \text{TeV})^2} ~,& \quad \frac{-1}{(2.3\, \text{TeV})^2} < \frac{C_{\mu\mu cc}^{RR}}{\Lambda^2} &< \frac{1}{(2.2\, \text{TeV})^2} ~.
  \end{align}
The above values are in good agreement with those we found in~\cite{Altmannshofer:2023bfk} calculating the cross sections without the \texttt{high-PT} package.

Assuming that the sensitivity to the high invariant mass di-lepton cross section scales with the square root of the luminosity, we can expect that the sensitivity to the new physics scale $\Lambda$ can improve by approximately a factor of 2 at the HL-LHC. 

\subsection{\texorpdfstring{\boldmath}{} Decays of the \texorpdfstring{$Z$}{Z} boson} \label{sec:Zdecay}

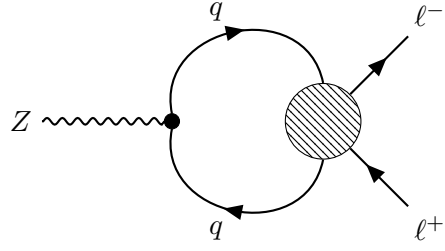
\begin{figure}[tb]
\centering
    \begin{tikzpicture}
        \begin{feynman}[large,every blob={/tikz/inner sep=0pt}]
            \vertex (Z) {\(Z\)};
            \node [right=of Z,style=dot] (a);
            \node[right=of a,style=blob] (b);
            \vertex[above right=of b] (f1) {\(\ell^{-}\)};
            \vertex[below right=of b] (f2) {\(\ell^{+}\)};
    
            \diagram*{
                (Z) -- [boson](a),
                (a) --[fermion, half left,edge label=\(q\)](b),
                (b) --[fermion, half left,edge label=\(q\)](a),
                (b) --[fermion](f1),
                (b)--[anti fermion](f2)
                
            };
        
        \end{feynman}
    \end{tikzpicture}
\caption{Example Feynman diagram for a 1-loop correction to the $Z$ decay into leptons. The fermion loop can be either a light quark or a top. Similar diagrams with lepton loops give corrections to $Z$ decays into quarks.} 
\label{fig:Zdecay}
\end{figure}

In the presence of the flavor-conserving operators, the decays of the $Z$ boson to quarks and leptons receive 1-loop corrections. 
An example diagram is shown in Figure~\ref{fig:Zdecay}. 

Our analysis of $Z$ boson decays follows closely the one outlined in~\cite{Altmannshofer:2023bfk}. Here we summarize the most important aspects. 
The corrections to the $Z$ decay rates can be written in the following way at the leading log approximation
\begin{multline}
    \label{eq:Zll}
    \frac{\Gamma(Z \to \ell \ell)}{\Gamma(Z\to \ell \ell)_\text{SM}} =
    1 + \frac{(1-2s_W^2) C^{LR}_{\ell\ell tt} -2s_W^2 C^{RR}_{\ell\ell tt}}
             {1-4s_W^2 + 8 s_W^4}
    \left(
        \frac{3}{4 \pi^2} \frac{m_t^2}{\Lambda^2}
        - \frac{s_W^2}{3\pi^2} \frac{m_Z^2}{\Lambda^2}
    \right)\log\left(\frac{\Lambda^2}{m_t^2}\right) \\
    + \frac{
        (2s_W^2-1)\left(C^{LR}_{\ell\ell uu}+C^{LR}_{\ell\ell cc}\right)
        + 2s_W^2\left(C^{RR}_{\ell\ell uu}+C^{RR}_{\ell\ell cc}\right)
    }{1-4s_W^2 + 8 s_W^4}
    \frac{s_W^2}{3\pi^2} \frac{m_Z^2}{\Lambda^2}
    \log\left(\frac{\Lambda^2}{m_Z^2}\right)~,
\end{multline}
\begin{multline}
    \label{eq:Znunu}
    \frac{\Gamma(Z \to \nu \nu)}{\Gamma(Z\to \nu \nu)_\text{SM}} =
    1 - \frac{1}{3}\Big(C^{LR}_{ee tt} + C^{LR}_{\mu\mu tt} \Big)
     \left(
        \frac{3}{4 \pi^2} \frac{m_t^2}{\Lambda^2}
        - \frac{s_W^2}{3\pi^2} \frac{m_Z^2}{\Lambda^2}
    \right) \log\left(\frac{\Lambda^2}{m_t^2}\right) \\
    + \frac{1}{3} \Big( C^{LR}_{ee uu} + C^{LR}_{\mu\mu uu}
        + C^{LR}_{ee cc} + C^{LR}_{\mu\mu cc} \Big)
     \frac{s_W^2}{3\pi^2} \frac{m_Z^2}{\Lambda^2}
    \log\left(\frac{\Lambda^2}{m_Z^2}\right)~,
\end{multline}
\begin{equation}
    \label{eq:Zqq}
    \frac{\Gamma(Z \to qq)}{\Gamma(Z\to qq)_\text{SM}} =
    1 + \frac{(2s_W^2-1)(C^{LR}_{ee qq}+C^{LR}_{\mu\mu qq}) + 2s_W^2(C^{RR}_{ee qq}+C^{RR}_{\mu\mu qq}) }{9-24s_W^2 +32 s_W^4}  \frac{s_W^2}{\pi^2} \frac{m_Z^2}{\Lambda^2} \log\left(\frac{\Lambda^2}{m_Z^2}\right)~,
\end{equation}
where $q = u,c $. The decay widths into down-type quarks remain SM-like. As in the expressions for the rare $B$ decays in section~\ref{sec:Bdecay}, $m_t \simeq 162.9$\,GeV is the top quark $\overline{\text{MS}}$ mass and $s_W^2 \simeq 0.231$ is the sine squared of the weak mixing angle. Similarly to the rare $B$ decays, we only include the logarithmically enhanced terms. Additional finite corrections have been worked out for example in~\cite{Dawson:2022bxd} and are of the same order as unknown model dependent UV matching contributions. The bounds we obtain from $Z$ decays thus hold barring cancellations with additional non-log terms. 

In our numerical analysis, we combine the SM predictions and experimental results given in~\cite{ALEPH:2005ab} into a $\chi^2$ function, taking into account the given correlations between the experimental uncertainties and neglecting the small uncertainties in the SM predictions. This allows us in principle to simultaneously constrain the various Wilson coefficients that enter equations~\eqref{eq:Zll}, \eqref{eq:Znunu}, and~\eqref{eq:Zqq}.

We demand that $\Delta \chi^2 = \chi^2 - \chi^2_\text{min} < 4$, where $\chi^2_\text{min}$ is the minimum $\chi^2$ obtainable if we let only one coefficient float.
Switching on one coefficient at a time, and fixing it to $\pm 1$, this procedure results in the following bounds on the new physics scale $\Lambda$
\begin{align} \label{eq:bounds_Zdecay_first}
  \frac{-1}{(0.24\, \text{TeV})^2} < \frac{C_{eeuu}^{LR}}{\Lambda^2} &< \frac{1}{(0.77\, \text{TeV})^2} ~,& \quad \frac{-1}{(0.62\, \text{TeV})^2} < \frac{C_{eeuu}^{RR}}{\Lambda^2} &< \frac{1}{(0.17\, \text{TeV})^2} ~, \\
  \frac{-1}{(0.24\, \text{TeV})^2} < \frac{C_{eecc}^{LR}}{\Lambda^2} &< \frac{1}{(0.77\, \text{TeV})^2} ~,& \quad \frac{-1}{(0.62\, \text{TeV})^2} < \frac{C_{eecc}^{RR}}{\Lambda^2} &< \frac{1}{(0.17\, \text{TeV})^2} ~, \\ \label{eq:bounds_Zdecay_third}
  \frac{-1}{(5.4\, \text{TeV})^2} < \frac{C_{eett}^{LR}}{\Lambda^2} &< \frac{1}{(2.2\, \text{TeV})^2} ~,& \quad \frac{-1}{(1.9\, \text{TeV})^2} < \frac{C_{eett}^{RR}}{\Lambda^2} &< \frac{1}{(4.5\, \text{TeV})^2} ~, \\
  \frac{-1}{(0.28\, \text{TeV})^2} < \frac{C_{\mu\mu uu}^{LR}}{\Lambda^2} & ~,& \quad  \frac{C_{\mu\mu uu}^{RR}}{\Lambda^2} &< \frac{1}{(0.35\, \text{TeV})^2} ~, \\
  \frac{-1}{(0.28\, \text{TeV})^2} < \frac{C_{\mu\mu cc}^{LR}}{\Lambda^2} & ~,& \quad  \frac{C_{\mu\mu cc}^{RR}}{\Lambda^2} &< \frac{1}{(0.35\, \text{TeV})^2} ~, \\
  \frac{-1}{(1.5\, \text{TeV})^2} < \frac{C_{\mu\mu tt}^{LR}}{\Lambda^2} &< \frac{1}{(2.4\, \text{TeV})^2} ~,& \quad \frac{-1}{(2.7\, \text{TeV})^2} < \frac{C_{\mu\mu tt}^{RR}}{\Lambda^2} &< \frac{1}{(1.2\, \text{TeV})^2} ~. \label{eq:bounds_Zdecay_last}
  \end{align}
The coefficients involving top quarks are most strongly constrained, because their contributions to the $Z$ decays are enhanced by the large top mass. The constraints on the other Wilson coefficients are rather weak, sometimes well below the 1 TeV scale. We do not find meaningful upper bounds in the case of $C^{LR}_{\mu\mu uu}$, $C^{LR}_{\mu\mu cc}$ and lower bounds in the case of $C^{RR}_{\mu\mu uu}$, $C^{RR}_{\mu\mu cc}$. 
Note that the above bounds continue to hold to good approximation if the values of the Wilson coefficients are varied by an order one amount. 

Future circular electron-positron colliders running on the $Z$-pole will produce more than $10^{12}$ $Z$ bosons and are expected to significantly improve the precision on the $Z$ decays. If systematic uncertainties can be controlled, FCC-ee and CEPC might improve the precision on the partial $Z$ decay widths by two orders of magnitude~\cite{CEPCStudyGroup:2018ghi, FCC:2018evy, CEPCPhysicsStudyGroup:2022uwl, Bernardi:2022hny}. This would correspond to an increase in sensitivity to the new physics scale of around one order of magnitude compared to what is shown in equations~\eqref{eq:bounds_Zdecay_first}-\eqref{eq:bounds_Zdecay_last}. (See also~\cite{Allwicher:2023shc, Knapen:2024bxw, Maura:2024zxz, Greljo:2024ytg} for related studies.)

The lepton flavor-violating operators can also induce the lepton flavor-violating decay $Z \to \mu e$ at 1-loop. However, existing limits and expected sensitivities on $Z \to \mu e$ from LEP~\cite{ALEPH:1991qhf, L3:1993dbo, OPAL:1995grn, DELPHI:1996iox}, the LHC~\cite{ATLAS:2022uhq} and future $Z$-pole machines~\cite{Dam:2018rfz, Altmannshofer:2022fvz} are not competitive with low energy constraints~\cite{Calibbi:2021pyh} and therefore we do not consider this decay in the following. 

\subsection{Top quark pair production in association with leptons at the LHC} \label{sec:ttbar}

\begin{figure}[tb]
\centering
    \begin{tikzpicture} 
        \begin{feynman}[large]
            \vertex [left=of a](g1) {\(g\)}; 
            \node [right=of g1,style=dot] (a); 
            \vertex [right=3.3cm of a](f1) {\(t\)};
            \node [below right=of a ,style=blob] (b);
            \node [below left=of b,style=dot](c);
            \vertex [right=of b,yshift=.7cm] (f2) {\(\ell^{-}\)};
            \vertex [right=of b,yshift=-.7cm] (f3) {\(\ell^{+}\)};
            \vertex[left=of c] (g2) {\(g\)};
            \vertex[right=3.3cm of c] (f4) {\(\bar{t}\)};
            
            \diagram* {
                (g1) -- [gluon] (a), 
                (a) -- [fermion] (f1),  
                (a) -- [anti fermion, edge label=\(t\)] (b),
                (b) -- [fermion] (f2),
                (b) -- [anti fermion] (f3),
                (b) -- [anti fermion,edge label=\(\bar{t}\) ] (c),
                (g2) -- [gluon] (c),
                (c) -- [anti fermion] (f4)
            }; 
        \end{feynman}
    \end{tikzpicture}
\caption{Example Feynman diagram for $t \bar t$ production in association with a lepton pair in proton-proton collisions.} 
\label{fig:ttll}
\end{figure}
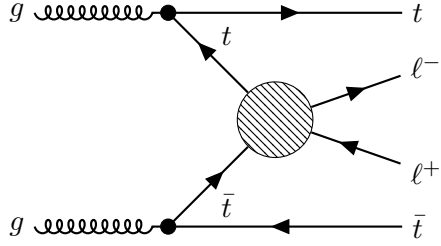

While the $Z$ and $B$ decays discussed above were affected by the flavor-conserving top quark operators at the 1-loop level, $t \bar t$ production in association with leptons at the LHC, $pp \to t\bar t \ell^+\ell^-$, receives tree-level contributions from such operators. One example diagram for a new physics contribution is shown in Figure~\ref{fig:ttll}. The corresponding bounds on the Wilson coefficients are more robust than the ones from $Z$ and $B$ decays, as they do not rely on a leading log approximation and the absence of cancelations with unrelated new physics contributions.

We use the results from the CMS search~\cite{CMS:2023xyc, CMS:2023xyc_suppl} to constrain the Wilson coefficients $C_{\ell\ell tt}^{LR}$ and $C_{\ell \ell tt}^{RR}$ (see also~\cite{CMS:2020lrr} for an earlier CMS search, and~\cite{ATLAS:2025yww} for a very recent ATLAS search). The CMS analysis directly presents constraints on the Wilson coefficients assuming lepton flavor universality $C_{ee tt}^{LR} = C_{\mu\mu tt}^{LR} = C_{\ell\ell tt}^{LR}$ and $C_{ee tt}^{RR} = C_{\mu\mu tt}^{RR} = C_{\ell\ell tt}^{RR}$. To recast the results for electron and muon-specific coefficients, we developed the following procedure:

The cross sections for $pp \to t\bar t e^+e^-$ and $pp \to t\bar t \mu^+\mu^-$ are quadratic functions of the respective Wilson coefficients. We assume that an experimental analysis would produce approximately Gaussian likelihoods for the Wilson coefficients entering each of the processes. The corresponding $\chi^2$ functions are then approximately described by fourth order polynomials. Switching on either the $LR$ or $RR$ coefficients, we have
\begin{eqnarray}
 \chi^2(C_{eett}^{XY}) &\simeq& \frac{a_e^{XY}}{2} + b_e^{XY} C_{eett}^{XY} + c_e^{XY} \left(C_{eett}^{XY}\right)^2 + d_e^{XY} \left(C_{eett}^{XY}\right)^3  + e_e^{XY} \left(C_{eett}^{XY}\right)^4 ~, \\
 \chi^2(C_{\mu\mu tt}^{XY}) &\simeq& \frac{a_\mu^{XY}}{2} + b_\mu^{XY} C_{\mu\mu tt}^{XY} + c_\mu^{XY} \left(C_{\mu\mu tt}^{XY}\right)^2 + d_\mu^{XY} \left(C_{\mu\mu tt}^{XY}\right)^3  + e_\mu^{XY} \left(C_{\mu\mu tt}^{XY}\right)^4 ~.
\end{eqnarray}
Assuming that the selection cuts and the detection efficiencies are approximately the same for $e^+ e^-$ and $\mu^+ \mu^-$ events, and assuming there are no significant statistical fluctuations, the electron coefficients and muon coefficients in these polynomials should be approximately the same.
The combined likelihood is then described by the following $\chi^2$ function
\begin{multline}\label{eq:chi2}
\chi^2(C_{eett}^{XY}, C_{\mu\mu tt}^{XY}) \simeq a^{XY} + b^{XY} \left( C_{eett}^{XY} + C_{\mu\mu tt}^{XY} \right) + c^{XY} \left( \left(C_{eett}^{XY}\right)^2+ \left(C_{\mu\mu tt}^{XY}\right)^2 \right) \\ + d^{XY} \left( \left(C_{eett}^{XY}\right)^3+\left(C_{\mu\mu tt}^{XY}\right)^3 \right)  + e^{XY} \left( \left(C_{eett}^{XY}\right)^4 +\left( C_{\mu\mu tt}^{XY}\right)^4 \right) ~.
\end{multline}
Assuming lepton flavor universality, as done in the CMS analysis, the $\chi^2$ function becomes
\begin{multline}\label{eq:chi2_CMS}
\chi^2(C_{\ell\ell tt}^{XY}) \simeq a^{XY} + 2 b^{XY} C_{\ell\ell tt}^{XY} + 2c^{XY} \left(C_{\ell\ell tt}^{XY}\right)^2 + 2 d^{XY} \left(C_{\ell\ell tt}^{XY}\right)^3 + 2 e^{XY} \left( C_{\ell\ell tt}^{XY}\right)^4 ~.
\end{multline}
We use the supplementary material provided on the CMS public page \cite{CMS:2023xyc_suppl}, that shows the log likelihood of the two lepton flavor universal Wilson coefficients $C_{\ell\ell tt}^{LR}$ and $C_{\ell\ell tt}^{RR}$. We extract the coefficients of the polynomials in equation~\eqref{eq:chi2_CMS} from the shown plots and apply them in the $\chi^2$ function~\eqref{eq:chi2}. 

Using this approach, we obtain the following constraints on the individual Wilson coefficients
\begin{eqnarray} \label{eq:bounds_tt_first_1}
  && \frac{-1}{{(0.60\, \text{TeV})^2}} < \frac{C_{ee tt}^{RR}}{\Lambda^2},~\frac{C_{\mu\mu tt}^{RR}}{\Lambda^2} <  \frac{1}{{(0.55\, \text{TeV})^2}} ~, \\ \label{eq:bounds_tt_first_2}
  && \frac{-1}{{(0.59\, \text{TeV})^2}} < \frac{C_{ee tt}^{LR}}{\Lambda^2},~\frac{C_{\mu\mu tt}^{LR}}{\Lambda^2} < \frac{1}{{(0.57\, \text{TeV})^2}} ~. 
\end{eqnarray}
We observe that these constraints are somewhat weaker than the ones we found from $B$ decays in equations~\eqref{eq:bound_B_tt_LR} and \eqref{eq:bound_B_tt_RR} and from $Z$ decays in equations~\eqref{eq:bounds_Zdecay_third} and~\eqref{eq:bounds_Zdecay_last}. However, as we argued above, these bounds are more robust.

With the expected increase in statistics from the HL-LHC, we naively conclude that the sensitivity of the $pp \to t\bar t \ell^+\ell^-$ measurements to the new physics scale can improve by approximately a factor of 2, reaching the TeV regime.

\subsection{Summary of the most sensitive probes} \label{sec:summary}

Before moving on to a detailed numerical study of the experimental constraints on the Wilson coefficients in the context of the sum rule relations shown in section~\ref{sec:positivity}, we briefly summarize the most sensitive probes of each set of coefficients (c.f. Figure~\ref{fig:chart}).

\bigskip\noindent
\underline{$C_{e\mu tq}^{LR}$, $C_{e\mu tq}^{RR}$, $C_{\mu etq}^{LR}$, $C_{\mu etq}^{RR}$}: The lepton flavor-violating and top flavor-violating coefficients are best probed by the LHC searches for rare top decays $t \to q e \mu$ and single top production in association with leptons $pp \to t e \mu$. The current constraints on the new physics scale shown in equations~\eqref{eq:tuemu_bound} and~\eqref{eq:tcemu_bound} are at the level of few TeV and are expected to improve by approximately $~50\%$, at the the HL-LHC. The constraints from rare $B$ decays are considerably weaker (around 300\,GeV, see equations (\ref{eq:C_rareB2}), (\ref{eq:C_rareB3})) and less robust, as they rely on a leading log approximation and the absence of cancellations with unrelated new physics. 

\bigskip\noindent
\underline{$C_{e\mu tt}^{LR}$, $C_{e\mu tt}^{RR}$, $C_{e\mu qq}^{LR}$, $C_{e\mu qq}^{RR}$}: The lepton flavor-violating and quark flavor-conserving coefficients are strongly constrained by bounds on $\mu$ to $e$ conversion in nuclei. As discussed in section~\ref{sec:lfv}, these constraints are so strong, that we can neglect the lepton flavor-violating and quark flavor-conserving operators in our analysis.

\bigskip\noindent
\underline{$C_{ee tq}^{LR}$, $C_{ee tq}^{RR}$, $C_{\mu\mu tq}^{LR}$, $C_{\mu\mu tq}^{RR}$}: The top flavor-violating but lepton flavor-conserving coefficients are fairly weakly constrained at the moment. In the case of electrons, the best constraints on the new physics scale come from single top production at LEP and are close to $1$\,TeV, see equation~\eqref{eq:bounds_single_top}. In the case of muons, the best constraints come from a recast of $t\to q Z$ searches and reach only a few hundred GeV, see equation~\eqref{eq:C_raretop2}. Rare $B$ decays give constraints in the same ballpark, but are less robust, see equation (\ref{eq:bound_B_tq_LR}).

\bigskip\noindent
\underline{$C_{ee qq}^{LR}$, $C_{ee qq}^{RR}$, $C_{\mu\mu qq}^{LR}$, $C_{\mu\mu qq}^{RR}$}: The flavor-conserving operators with light quarks are best probed by di-lepton production at the LHC. We find that the constraints on the new physics scale of up quark operators are approaching 10~TeV, see equations~\eqref{eq:bounds_ppll_1} and~\eqref{eq:bounds_ppll_3}, while those of charm quark operators are around 2~TeV, see equations~\eqref{eq:bounds_ppll_2} and~\eqref{eq:bounds_ppll_4}. The sensitivity to the new physics scale can improve by a factor of approximately~2 at the HL-LHC. Constraints from precision measurements of $Z$ decays at LEP are much weaker (around a few hundred GeV) and also less robust, as they are based on a leading log approximation and assume the
absence of cancellations with other possible new physics contributions.

\bigskip\noindent
\underline{$C_{ee tt}^{LR}$, $C_{ee tt}^{RR}$, $C_{\mu\mu tt}^{LR}$, $C_{\mu\mu tt}^{RR}$}: The flavor-conserving operators with top quarks are bounded because of their loop level contributions to rare $B$ decays and leptonic $Z$ decays. The corresponding constraints on the new physics scale are around a few TeV, with the precise values collected in equations~\eqref{eq:bound_B_tt_LR} and~\eqref{eq:bound_B_tt_RR} (rare $B$ decays) and in equations~\eqref{eq:bounds_Zdecay_third} and~\eqref{eq:bounds_Zdecay_last} ($Z$ decays). As already emphasized, the $B$ decay and $Z$ decay constraints only hold barring accidental cancellations with other contributions. This is in contrast to the constraints that can be derived from $t \bar t$ production in association with leptons. The latter constraints are weaker (below the TeV scale, see equations~\eqref{eq:bounds_tt_first_1} and~\eqref{eq:bounds_tt_first_2}) but robust. 

\begin{figure}
    \centering
    \begin{subfigure}[t]{1\textwidth}
        \centering
        \includegraphics[width=\linewidth]{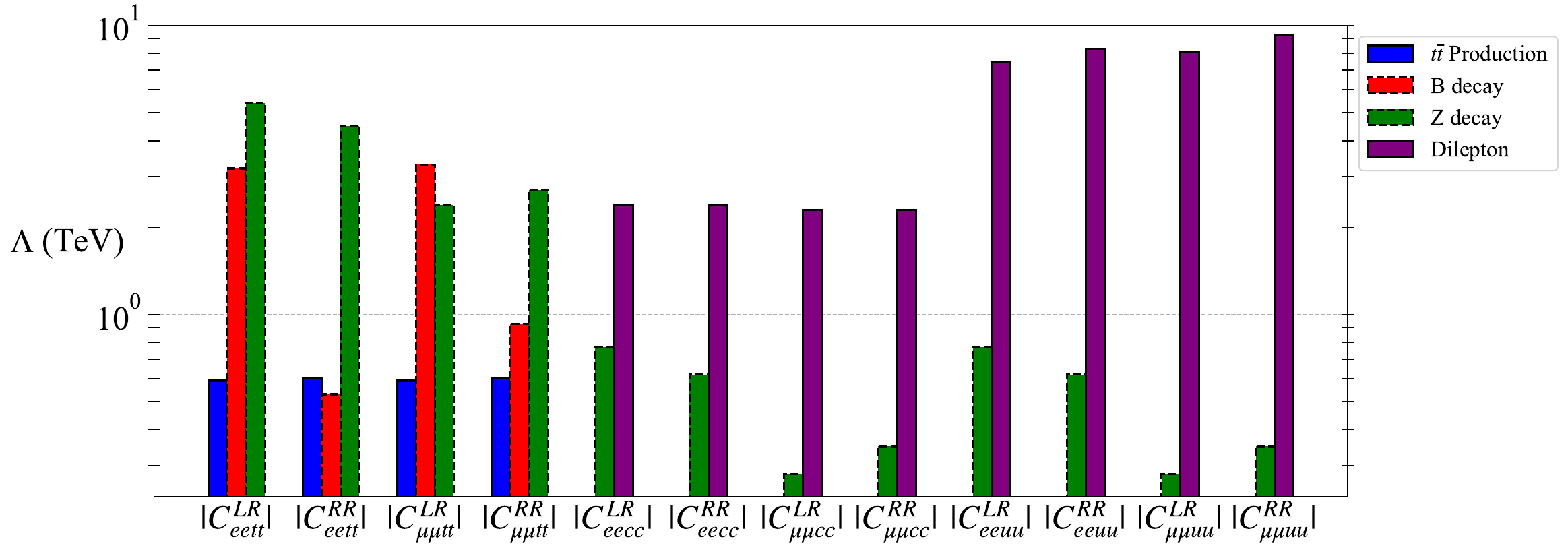}
    \end{subfigure}

    \begin{subfigure}[t]{1\textwidth}
        \centering
        \includegraphics[width=\linewidth]{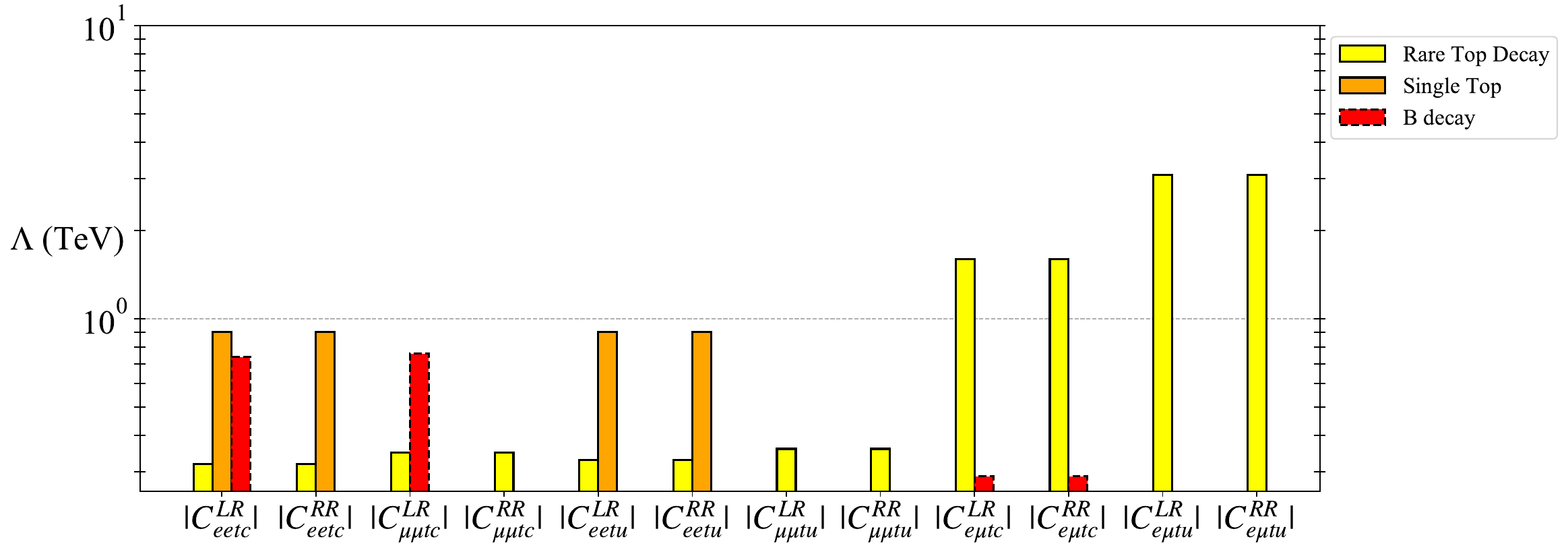}
    \end{subfigure}  
    \caption{Bar charts illustrating the various constraints on the Wilson coefficients. The vertical axis shows the new physics scale $\Lambda$ that is probed, switching on a single Wilson coefficient at a time with absolute value $1$. In cases for which the constraints for Wilson coefficient values of $+1$ and $-1$ differ, we show the stronger of the two. The upper (lower) plot shows flavor-conserving (flavor-violating) Wilson coefficients. The constraints on the $C_{\mu e tq}^{XY}$ coefficients are the same as the ones on the $C_{e \mu tq}^{XY}$ coefficients and not shown. }  \label{fig:bar_charts}
\end{figure}

\bigskip\noindent
All the constraints summarized in this section are also illustrated in the plots of Figure~\ref{fig:bar_charts}. 

\section{Numerical Analysis} \label{sec:numerics}

Having outlined the sum rule relations for the Wilson coefficients in section~\ref{sec:positivity} and the relevant experimental constraints in section~\ref{sec:observables}, we will now turn this information into indirect bounds on branching ratios of rare top decays.  
 
We assume that all Wilson coefficients are real. Including also imaginary parts of the Wilson coefficients might have a non-trivial impact on observables in which the new physics interferes with the SM amplitudes. As the SM amplitudes of the processes we consider are real to an excellent approximation, only the real part of the new physics Wilson coefficient is able to interfere, while imaginary parts enter the observables at the quadratic level. This implies that some experimental constraints might get softened in the presence of imaginary parts of the Wilson coefficients. A detailed study of the impact of complex Wilson coefficients is left for future work.

We start our discussion by focusing on lepton flavor-conserving rare top decays in section~\ref{sec:numerics_DF1}. This section updates the results presented in~\cite{Altmannshofer:2023bfk}. In section~\ref{sec:numerics_DF2} we cover lepton flavor-violating rare top decays, which is a novel aspect.  

\subsection{Lepton flavor-conserving rare top decays} \label{sec:numerics_DF1}

We consider eight different scenarios with three non-zero Wilson coefficients each. These are minimal scenarios that allow us to illustrate the impact of the Wilson coefficient relations from equation~\eqref{eq:sum_rule_DeltaF1}.
\begin{eqnarray}
\text{``electron-up LR'':} && \quad C_{eeuu}^{LR} ~,~~ C_{eett}^{LR} ~,~~ C_{ee tu}^{LR} ~, \\
\text{``electron-charm LR'':} && \quad C_{eecc}^{LR} ~,~~ C_{eett}^{LR} ~,~~ C_{ee tc}^{LR} ~, \\
\text{``electron-up RR'':} && \quad C_{eeuu}^{RR} ~,~~ C_{eett}^{RR} ~,~~ C_{ee tu}^{RR} ~, \\
\text{``electron-charm RR'':} && \quad C_{eecc}^{RR} ~,~~ C_{eett}^{RR} ~,~~ C_{ee tc}^{RR} ~, \\
\text{``muon-up LR'':} && \quad C_{\mu\mu uu}^{LR} ~,~~ C_{\mu\mu tt}^{LR} ~,~~ C_{\mu\mu tu}^{LR} ~, \\
\text{``muon-charm LR'':} && \quad C_{\mu\mu cc}^{LR} ~,~~ C_{\mu\mu tt}^{LR} ~,~~ C_{\mu\mu tc}^{LR} ~, \\
\text{``muon-up RR'':} && \quad C_{\mu\mu uu}^{RR} ~,~~ C_{\mu\mu tt}^{RR} ~,~~ C_{\mu\mu tu}^{RR} ~, \\
\text{``muon-charm RR'':} && \quad C_{\mu\mu cc}^{RR} ~,~~ C_{\mu\mu tt}^{RR} ~,~~ C_{\mu\mu tc}^{RR} ~.
\end{eqnarray}
Switching on more Wilson coefficients simultaneously, may soften some of the constraints if one is willing to consider cancellations in the new physics contributions to the observables we consider.

In each of the above scenarios we numerically determine the maximal value for the rare top branching ratios that is compatible with the relations given in equation~\eqref{eq:sum_rule_DeltaF1}, and with the flavor-conserving Wilson coefficients subject to the relevant constraints discussed in section~\ref{sec:observables}. Note that this procedure goes beyond combining the constraints we have summarized for example in Figure~\ref{fig:bar_charts}, as now several Wilson coefficients are nonzero simultaneously, thus requiring a re-evaluation of many of the experimental constraints. We find
\begin{eqnarray} \label{eq:BR_limit_first}
\text{``electron-up LR'':} && \quad  \text{BR}(t \to u e^+ e^-) < 1.8 \times 10^{-8} ~~ (1.6 \times 10^{-7}) ~, \\\label{eq:BR_limit_second}
\text{``electron-charm LR'':} && \quad  \text{BR}(t \to c e^+ e^-) < 1.3 \times 10^{-7} ~~ (1.1 \times 10^{-6}) ~, \\
\text{``electron-up RR'':} && \quad  \text{BR}(t \to u e^+ e^-) <1.2 \times 10^{-8} ~~ (2.1 \times 10^{-7}) ~, \\
\text{``electron-charm RR'':} && \quad  \text{BR}(t \to c e^+ e^-) < 1.5 \times 10^{-7} ~~ (1.8 \times 10^{-6}) ~, \\
\text{``muon-up LR'':} && \quad  \text{BR}(t \to u \mu^+ \mu^-) < 1.9 \times 10^{-8} ~~ (1.8 \times 10^{-7})~, \\
\text{``muon-charm LR'':} && \quad  \text{BR}(t \to c \mu^+ \mu^-) < 2.4 \times 10^{-7} ~~ (1.3 \times 10^{-6}) ~, \\
\text{``muon-up RR'':} && \quad  \text{BR}(t \to u \mu^+ \mu^-) < 6.7 \times 10^{-8} ~~ (2.4 \times 10^{-7})~, \\
\text{``muon-charm RR'':} && \quad  \text{BR}(t \to c \mu^+ \mu^-) < 3.7 \times 10^{-7} ~~ (1.4 \times 10^{-6}) ~. \label{eq:BR_limit_last}
\end{eqnarray}
For the first values quoted above, we take into account all constraints discussed in section~\ref{sec:observables}. In that case, we find upper limits on branching ratios at the level of $10^{-7}$ (for the decays into charm quarks) and $10^{-8}$ (for the decays into up quarks). To get the values in parentheses, we exclude the $Z$ and $B$ decay constraints and rely on the $t\bar t$ production in association with leptons to constrain the flavor-conserving operators with top quarks. As the corresponding bounds on the Wilson coefficients are weaker by approximately an order of magnitude (see Figure \ref{fig:bar_charts}), we obtain approximately an order of magnitude weaker bounds on the rare top decay branching ratios.

\begin{figure}[tb]
\centering
\includegraphics[width=1.0\textwidth]{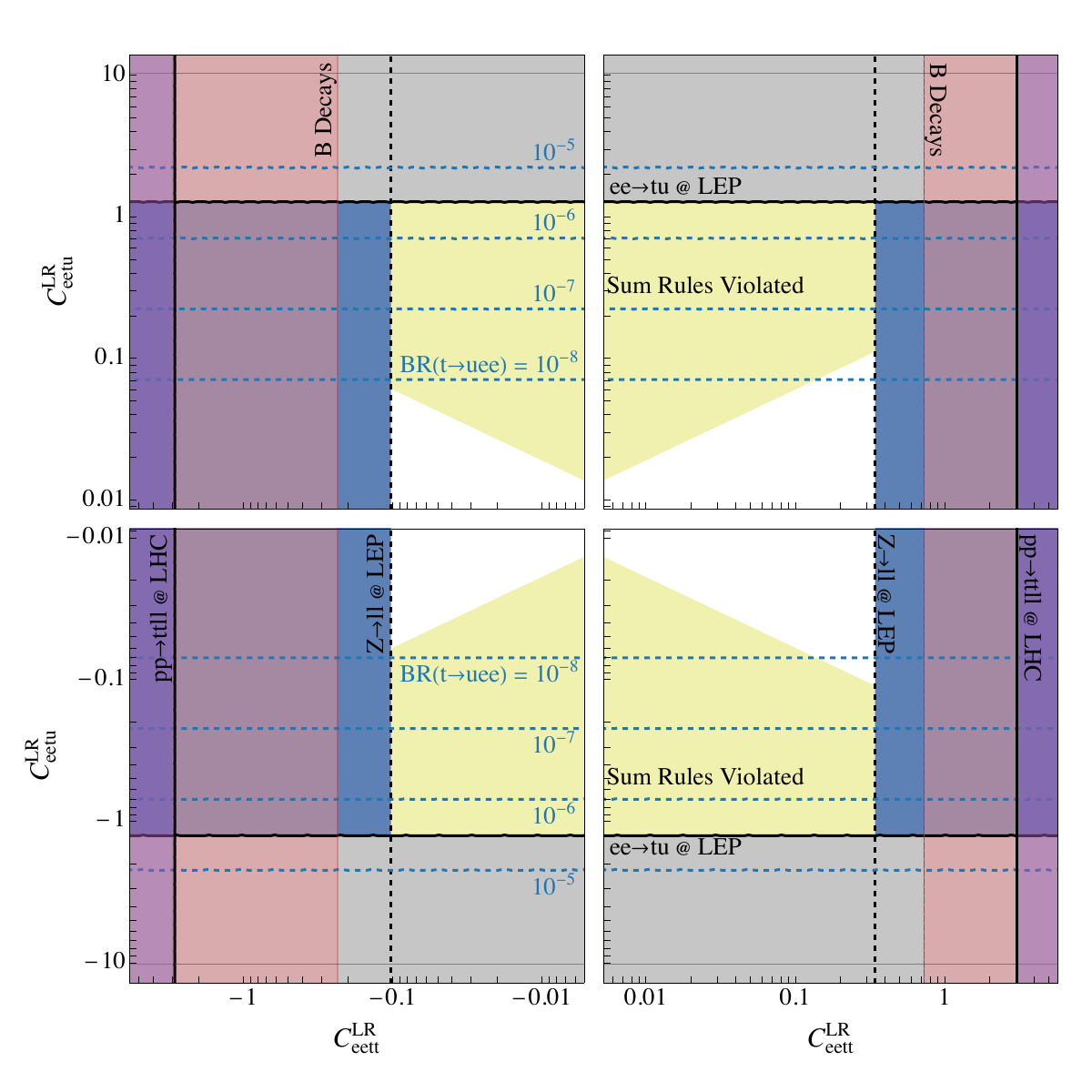}
\caption{Expected $t \to u e^+e^-$ branching ratios (dashed blue lines) in the ``electron-up LR'' scenario as function of the Wilson coefficients $C_{ee tt}^{LR}$ and $C_{ee tu}^{LR}$. The regions shaded in blue, red, and purple are excluded at the 95\% C.L. by constraints from $Z$ decays, $B$ decays, and $t \bar t$ production in association with leptons, respectively. The light gray region shows 95\% C.L. limits from single top production at LEP. In the yellow regions of parameter space, the ``sum rule'' relations of equation~\eqref{eq:sum_rule_DeltaF1} are violated. }
\label{fig:CeeutLR}
\end{figure}
\begin{figure}[tb]
\centering
\includegraphics[width=1.0\textwidth]{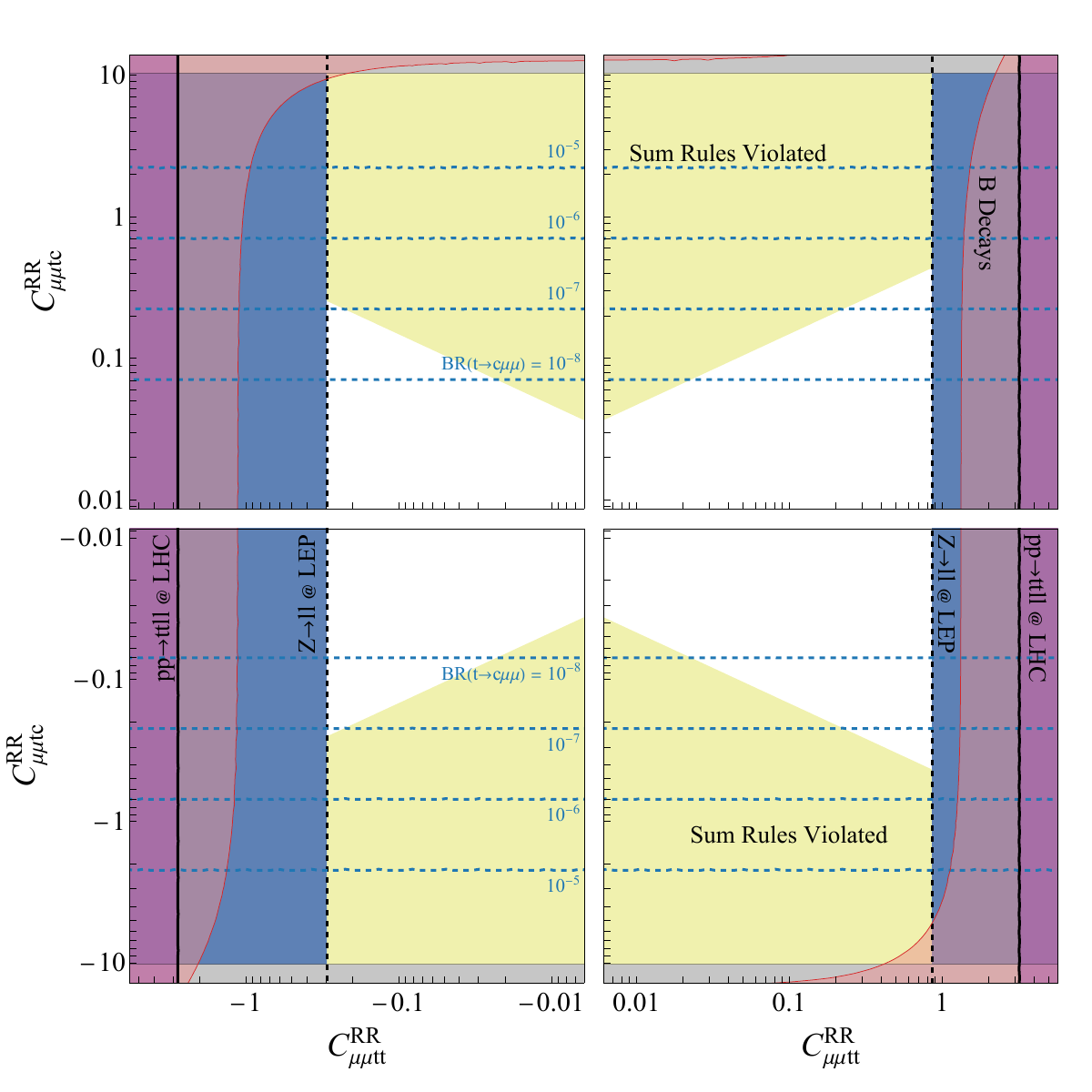}
\caption{Expected $t \to c \mu^+\mu^-$ branching ratios (dashed blue lines) in the ``muon-charm RR'' scenario as function of the Wilson coefficients $C_{\mu\mu tt}^{RR}$ and $C_{\mu\mu tc}^{RR}$. The regions shaded in blue, red, and purple are excluded at the 95\% C.L. by constraints from $Z$ decays, $B$ decays, and $t \bar t$ production in association with leptons, respectively. The light gray region shows 95\% C.L. limits from single top production at LEP. In the yellow regions of parameter space, the ``sum rule'' relations of equation~\eqref{eq:sum_rule_DeltaF1} are violated.}
\label{fig:CmmctRR}
\end{figure}
\begin{figure}[tb]
\centering
\includegraphics[width=1.0\textwidth]{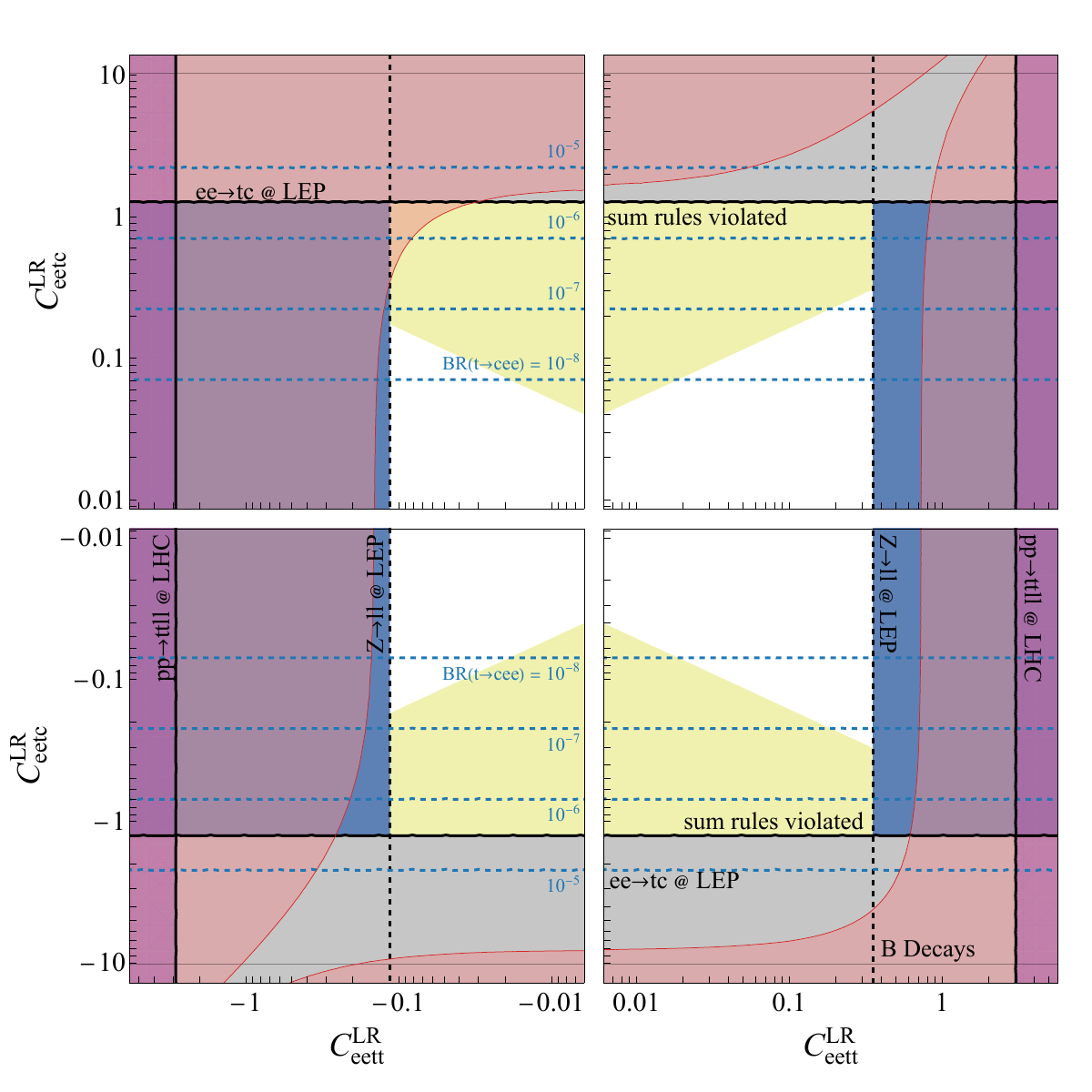}
\caption{Expected $t \to c e^+e^-$ branching ratios (dashed blue lines) in the ``electron-charm LR'' scenario as function of the Wilson coefficients $C_{ee tt}^{LR}$ and $C_{ee tc}^{LR}$. The regions shaded in blue, red, and purple are excluded at the 95\% C.L. by constraints from $Z$ decays, $B$ decays, and $t \bar t$ production in association with leptons, respectively. The light gray region shows 95\% C.L. limits from single top production at LEP. In the yellow regions of parameter space, the ``sum rule'' relations of equation~\eqref{eq:sum_rule_DeltaF1} are violated.}
\label{fig:CeectLR}
\end{figure}

The above limits on the rare top branching ratios are also illustrated in the plots of Figures~\ref{fig:CeeutLR}, \ref{fig:CmmctRR}, and \ref{fig:CeectLR} corresponding to the three representative example scenarios ``electron-up LR'' (equation (\ref{eq:BR_limit_first})),  ``muon-charm RR'' (equation (\ref{eq:BR_limit_last})), and ``electron-charm LR'' (equation (\ref{eq:BR_limit_second})), respectively. In each case, we show the relevant flavor-conserving Wilson coefficient with top quarks on the horizontal axis and the relevant top flavor-violating one on the vertical axis. The relevant flavor-conserving coefficient with light quarks is set to its maximal value, compatible with experimental constraints. Constraints from $Z$ decays, $B$ decays, and $t\bar t$ production in association with leptons are shown by the shaded regions in blue, red, and purple. Where applicable, the constraint from single top production at LEP is shown in gray.
Expected values for the rare top decays are shown by the dashed blue contours. In the yellow regions, the sum rule relations from equation~\eqref{eq:sum_rule_DeltaF1} are violated. The values in equations~\eqref{eq:BR_limit_first} - \eqref{eq:BR_limit_last} correspond to the maximal rare top decay branching ratios in the white regions of the plots. 
 
We stress that the values summarized above in equations~\eqref{eq:BR_limit_first} - \eqref{eq:BR_limit_last}, and the yellow regions in the plots of Figures~\ref{fig:CeeutLR}, \ref{fig:CmmctRR}, and \ref{fig:CeectLR} are not strict bounds, as they only hold in some UV models but can be violated in others (c.f the discussion at the end of section~\ref{sec:positivity}). Therefore, the bounds should rather be interpreted as sensitivity targets. Note that~\cite{Chala:2018agk} estimates that branching ratios at the level of $10^{-6}$ are within reach of the HL-LHC. If future searches for rare top decays find indeed a signal with a branching ratio above the values in parentheses, we will be able to conclude that the effect is due to a class of new physics that is not subject to the positivity considerations leading to the relations in equation~\eqref{eq:sum_rule_DeltaF1}. Models with $Z^\prime$ bosons would be the most prominent candidates. An observation of rare top decays with branching ratios between the two numbers would either point to the same class of new physics or might indicate that there is additional new physics in $Z$ decays and rare $B$ decays that renders their constraints on top operators ineffective.

We note that the sensitivity targets will change with more precise LHC measurements of di-lepton production and $t \bar t$ production in association with leptons. If better bounds on new physics in these processes can be established in the future, the target region for the rare top decays will move to smaller values. 

\subsection{Lepton flavor-violating rare top decays} \label{sec:numerics_DF2}

We focus on new physics that affects either up quarks or charm quarks but not both, which can be motivated by appropriate flavor symmetries. Moreover, we consider new physics either in left-handed or right-handed leptons, leading to the following four different scenarios with six nonzero Wilson coefficients each:
 
\begin{eqnarray}
\text{``up LR'':} && \quad C_{eeuu}^{LR} ~,~~ C_{\mu\mu uu}^{LR} ~,~~ C_{eett}^{LR} ~,~~ C_{\mu\mu tt}^{LR} ~,~~ C_{e\mu tu}^{LR} ~,~~ C_{\mu e tu}^{LR} ~, \\
\text{``charm LR'':} && \quad C_{eecc}^{LR} ~,~~ C_{\mu\mu cc}^{LR} ~,~~ C_{eett}^{LR} ~,~~ C_{\mu\mu tt}^{LR} ~,~~ C_{e\mu tc}^{LR} ~,~~ C_{\mu e tc}^{LR} ~, \\
\text{``up RR'':} && \quad C_{eeuu}^{RR} ~,~~ C_{\mu\mu uu}^{RR} ~,~~ C_{eett}^{RR} ~,~~ C_{\mu\mu tt}^{RR} ~,~~ C_{e\mu tu}^{RR} ~,~~ C_{\mu e tu}^{RR} ~, \\
\text{``charm RR'':} && \quad C_{eecc}^{RR} ~,~~ C_{\mu\mu cc}^{RR} ~,~~ C_{eett}^{RR} ~,~~ C_{\mu\mu tt}^{RR} ~,~~ C_{e\mu tc}^{RR} ~,~~ C_{\mu e tc}^{RR} ~.
\end{eqnarray}

Analogously to what we did in the previous section, we numerically determine the maximal value for the rare top branching ratios taking into account the sum rule relations for the $\Delta F = 2$ Wilson coefficients in equation~\eqref{eq:sum_rule_DeltaF2} and the experimental constraints on the flavor-conserving Wilson coefficients. We find
\begin{eqnarray}
\text{``up LR'':} && \quad \text{BR}(t \to u \mu e) < 1.2 \times 10^{-8} ~~   (3.5\times 10^{-7})~, \\
\text{``charm LR'':} && \quad\text{BR}(t \to c \mu e) < 9.0 \times 10^{-8} ~~  (3.9\times 10^{-7})~, \\
\text{``up RR'':} && \quad \text{BR}(t \to u \mu e) < 2.0 \times 10^{-8} ~~  (3.0\times 10^{-7})~, \\
\text{``charm RR'':} && \quad \text{BR}(t \to c \mu e) < 2.9 \times 10^{-7} ~~  (4.1\times 10^{-6})~.
\end{eqnarray}
%
%
where the first values take into account the bounds from $Z$ decays and rare $B$ decays, while the values in parentheses do not make use of these observables, but rely on $t \bar t$ production in association with leptons to constrain the flavor-conserving Wilson coefficients with top quarks. We find upper limits that are comparable to the case of lepton flavor-conserving decays.


%
%
We also stress here that the quoted values should be interpreted as targets for experimental searches. They hold in leptoquark models but can be violated in other new physics scenarios, in particular $Z^\prime$ models. Interestingly enough, these targets are in the same ballpark as the existing limits from searches at the LHC, see e.g. equation~\eqref{eq:bounds_tqllp}. This means that the next round of updates for $t \to q \mu e$ searches at the LHC will have reach beyond the ``sum rule targets'' and thus be sensitive to a wide range of new-physics scenarios.  

\section{Conclusions} \label{sec:conclusions}

In certain classes of new physics scenarios, there are relations between the Wilson coefficients of flavor-conserving and flavor-violating four-fermion operators. 
We have explored these relations in the context of top flavor-violating processes. The relations allow us to distinguish experimentally between different UV models, as they hold in leptoquark models but can be violated, for example, in $Z^\prime$ models.

On the one hand, we have updated our previous analysis in~\cite{Altmannshofer:2023bfk} and provided target regions for LHC searches for the rare top decays $t \to c \mu^+ \mu^-$, $t \to u \mu^+ \mu^-$, $t \to c e^+ e^-$, $t \to u e^+ e^-$. Taking into account existing experimental constraints from flavor-conserving processes, in particular di-lepton production at the LHC and $t \bar t$ production in association with leptons, we find interesting target regions for the rare top decay branching ratios in the ballpark of $10^{-7}$ to $10^{-6}$. Observation of those rare top decays with larger rates would point to specific new physics scenarios that are not subject to the theoretical relations. 

On the other hand, we have extended our analysis and also considered lepton flavor-violating decays of the top quark, $t \to c \mu e$ and $t \to u \mu e$. In a first step, we presented the non-trivial theoretical relations that hold between the relevant $\Delta F = 2$ four-fermion contact interactions and the flavor-conserving interactions. In a second step, we have confronted these relations with the current experimental information and found that present LHC searches for lepton flavor-violating top decays have already a sensitivity comparable to the target regions. This implies that future improved searches for $t \to q \mu e$ decays at the LHC will probe a wide range of new physics scenarios. 

Our work provides motivation to continue searches for top flavor-violating processes at the LHC and establishes interesting target sensitivities. In our analysis, we have focused on processes with muons and electrons. However, similar studies could also be performed for processes involving taus. Such studies would provide complementary information on the connection of all lepton flavors in rare top decays.

\section*{Acknowledgements}

We thank Lorenzo Calibbi for useful discussions.
The Feynman diagrams for this paper were drawn using \texttt{tikz-feynman}~\cite{Ellis:2016jkw}.
The research of WA, SG, and CMD is supported by the U.S. Department of Energy grant number DE-SC0010107. This research was supported in part by grant NSF PHY-2309135 to the Kavli Institute for Theoretical Physics (KITP) and was performed in part at the Aspen Center for Physics, which is supported by National Science Foundation grant PHY-2210452.

\bibliographystyle{JHEP.bst}
\bibliography{bibliography}

\end{document}